\documentclass{aa}

\usepackage{blindtext}
\usepackage{graphicx}
\usepackage{multicol}
\usepackage{tabularx}
\usepackage{rotating, booktabs}
\usepackage{txfonts}
\usepackage{filecontents}
\usepackage{threeparttable}
\usepackage{hyperref}
\usepackage{natbib}
\usepackage[toc,page]{appendix}
\usepackage{subfig}
\usepackage{lscape}
\bibpunct{(}{)}{;}{a}{}{,}

\usepackage{gensymb}

\def\ltsima{$\; \buildrel < \over \sim \;$}
\def\simlt{\lower.5ex\hbox{\ltsima}}
\def\gtsima{$\; \buildrel > \over \sim \;$}
\def\simgt{\lower.5ex\hbox{\gtsima}}

\def\ergs{{erg s$^{-1}$}}
\def\cm2{{cm$^{-2}$}}

\def\edd{$\lambda_\textup{Edd}$}

\begin{document}

\title{The Type II AGN-host galaxy connection: insights from the VVDS and VIPERS surveys}
  
\date{}
\author{{G. Vietri}\inst{\ref{inst1}}
\and
{B. Garilli}\inst{\ref{inst1}}
\and
{M. Polletta}\inst{\ref{inst1}} 
\and
{S. Bisogni} \inst{\ref{inst1}}
\and
{L. P. Cassarà}\inst{\ref{inst1}}
\and
{P. Franzetti}\inst{\ref{inst1}}
\and
{M. Fumana}\inst{\ref{inst1}}
\and
{A. Gargiulo}\inst{\ref{inst1}}
\and
{D. Maccagni}\inst{\ref{inst1}}
\and
{C. Mancini}\inst{\ref{inst1}}
\and
{M. Scodeggio}\inst{\ref{inst1}}
\and
{A. Fritz}\inst{\ref{inst2}}
\and
{K. Malek}\inst{\ref{inst3},\ref{inst4}}
\and
{G. Manzoni}\inst{\ref{inst5},\ref{inst6},\ref{inst7}}
\and
{A. Pollo}\inst{\ref{inst3},\ref{inst8}}
\and
{M. Siudek}\inst{\ref{inst9},\ref{inst10}}
\and
{D. Vergani}\inst{\ref{inst11}}
\and
{G. Zamorani}\inst{\ref{inst11}}
\and
{A. Zanichelli}\inst{\ref{inst12}}
}
    \institute{INAF -- Istituto di Astrofisica Spaziale e Fisica Cosmica di Milano, Via A. Corti 12, I-20133 Milano, Italy. \\ \email{giustina.vietri@inaf.it}\label{inst1}
    \and
    OmegaLambdaTec GmbH, Lichtenbergstra{\ss}e 8, D-85748 Garching, Germany\label{inst2}
    \and
    National Centre for Nuclear Research, ul. Hoza 69, PL-00-681 Warsaw, Poland\label{inst3}
    \and
    Aix Marseille Univ., CNRS, CNES, LAM Marseille, France \label{inst4}
    \and
    Institute for Computational Cosmology (ICC), Department of Physics, Durham University, South Road, Durham DH1 3LE, UK\label{inst5}
    \and
    Centre for Extragalactic Astronomy (CEA), Department of Physics, Durham University, South Road, Durham DH1 3LE, UK\label{inst6}
    \and
     Institute for Data Science (IDAS), Durham University, South Road, Durham DH1 3LE, UK\label{inst7}
     \and
     Astronomical Observatory of the Jagiellonian University, ul. Orla 171, PL-30-244 Kraków, Poland\label{inst8}
     \and
     Institut de F\'{\i}sica d'Altes Energies (IFAE), The Barcelona Institute of Science and Technology, 08193 Bellaterra (Barcelona), Spain\label{inst9}
     \and
     National Centre for Nuclear Research, ul. Pasteura 7, 02-093, Warsaw, Poland\label{inst10}
    \and
    INAF - Istituto di Astrofisica Spaziale e Fisica Cosmica Bologna, via Gobetti 101, I-40129 Bologna, Italy\label{inst11}
    \and
    INAF - Istituto di Radioastronomia, Via Gobetti 101, 40129, Bologna, Italy\label{inst12}}
 
  \abstract
 {We present a study of optically-selected Type II AGN at 0.5 < z < 0.9 from the VIPERS and VVDS surveys, to investigate the connection between AGN activity and physical properties of their host galaxies. The host stellar mass is estimated through spectral energy distribution fitting with the CIGALE code, and star formation rates are derived from the [OII]$\lambda$3727 \AA\ line luminosity.  We find that 49\% of the AGN host galaxies are on or above the main sequence (MS), 40\% lie in the sub-MS locus, and 11\% in the quiescent locus. Using the [OIII]$\lambda$5007 \AA\ line luminosity as a proxy of the AGN power, we find that at fixed AGN power Type II AGN host galaxies show a bimodal behaviour: systems with host galaxy stellar mass <10$^{10}$ M$_{\odot}$, reside along the MS or in the starbursts locus (high-SF Type II AGN), while systems residing in massive host-galaxies (>10$^{10}$ M$_{\odot}$) show a lower level of star formation (low-SF Type II AGN).
At all stellar masses, the offset from the MS is positively correlated with the AGN power. We interpret this correlation as evidence of co-evolution between the AGN and the host, possibly due to the availability of cold gas. In the most powerful AGN with host galaxies below the MS we find a hint, though weak, of asymmetry in the [OIII] line profile, likely due to outflowing gas, consistent with a scenario in which AGN feedback removes the available gas and halts the star formation in the most massive hosts.
 

}
   \keywords{galaxies: active – galaxies: nuclei – quasars: emission lines - quasars: general}

   \maketitle
%
%

\section{Introduction}\label{sec:intro}

Galaxies can be divided into two main types, blue star-forming galaxies that typically have disky morphologies, and red passive galaxies that on the contrary are bulge-dominated (e.g. \citealt{Gadotti2009}; \citealt{Bluck2014}; \citealt{Whitaker2015}). These two types can be easily identified in a color-magnitude diagram where they form, respectively, the blue cloud and the red sequence (e.g. \citealt{Strateva2001}; \citealt{Blanton2003}; \citealt{Kauffmann2003b}; \citealt{Baldry2004}). According to the current theory of galaxy evolution, galaxies move from the blue cloud to the red sequence (\citealt{Cowie1996}; \citealt{Baldry2004}; \citealt{Perez2008}, \citealt{Fritz2014}). The mechanisms involved in such a transformation, called galaxy quenching, are still a matter of study as they have to explain how the morphological transformation occurs, how star formation ceases, whether the environment and a galaxy stellar mass play a role, and when and on which timescale such a process occurs. Galaxy evolution models reproduce quenching by shutting off the cold gas supply in a galaxy (e.g \citealt{Gabor2010}). This can occur by inhibiting the cold gas from entering a galaxy or from producing stars, or through ejective feedback mechanisms that remove gas from the galaxy. One of the most viable ejective mechanisms is provided by powerful active galactic nuclei (AGN), powered by accretion onto supermassive black holes (SMBH, \citealt{Lynden1969}).

The discovery that black hole (BH) masses of nearby bulges correlate with the stellar velocity dispersion, mass and luminosity of the bulge (e.g. \citealt{Magorrian1998}, \citealt{Ferrarese2000}, \citealt{Gebhardt2000}, \citealt{Gultekin2009}, \citealt{Kormendy2013}), and the similarities between the evolution of the star formation rate density and the growth of the AGN (e.g. \citealt{Madau1996}, \citealt{Hasinger2005}, \citealt{Hopkins2007}, \citealt{Aird2015}) led to the idea that SMBH and their host-galaxies are tightly linked, despite the different size scales involved. 
Most galaxies have gone through an active phase during which the SMBH have accreted material, grown itself and possibly supplied the energy to influence the host galaxy on large scale distances (e.g. \citealt{Cattaneo2009}, \citealt{Kauffmann2000}). 
This could be possible through large-scale outflows expelling a large fraction of the gas from the host-galaxy (e.g. \citealt{Silk1998}), where a small fraction of the energy released by the BH accretion would be sufficient to heat and blow out the host-galaxy gas content. By including AGN feedback in numerical simulations and semi-analytical models of galaxy evolution a good agreement with the observations has been obtained, such as suppression of star formation at the highest stellar masses, which appears necessary to recover the properties of the local galaxy population (e.g. \citealt{DiMatteo2005}, \citealt{Croton2006}, \citealt{Schaye2015}, \citealt{Manzoni2021}). However, the impact AGN might have on their hosts and on the star formation activity is still a matter of numerous investigations (e.g. \citealt{Alexander2012}, \citealt{Kormendy2013}).

Previous works have studied the link between the AGN and their host galaxies, but with conflicting results. Some find that the strength of the AGN activity strongly correlates with the SFR of their host (e.g. \citealt{Mullaney2012}, \citealt{Chen2013}, \citealt{Hickox2014}, \citealt{Lanzuisi2017}, \citealt{Stemo2020}, \citealt{Zhuang2020}), whereas others find that SFR is weakly or not correlated with the AGN luminosity (e.g. \citealt{Azadi2015}, \citealt{Stanley2015,Stanley2017}, \citealt{Shimizu2017}). A dependence on redshift and luminosity seems to exist, with higher luminosity AGN (L$_{AGN}$ > 10$^{44}$ erg s$^{-1}$) and lower redshift (z < 1) galaxies exhibiting a steep correlation, while  no correlation is found for lower luminosities or AGN at higher redshifts (e.g., \citealt{Shao2010}; \citealt{Harrison2012}; \citealt{Rosario2012}, \citealt{Santini2012}). These inconclusive results could be due to different binning methods, e.g. AGN luminosity is averaged in bins of host properties as SFR and stellar mass, or the SFR is averaged in bins of AGN luminosity. As \cite{Hickox2014} point out, the different results likely arise from AGN luminosity varying on timescales shorter than that of SFR. Other factors include the sample size (e.g. \citealt{Harrison2012}, \citealt{Page2012}), selection effects, low number statistics, SFR measurements and the mutual dependence of AGN luminosity and SFR on stellar mass (\citealt{Harrison2017}).

Most star-forming galaxies show a tight correlation between the star formation rate (SFR) and the stellar mass, referred to as the Main Sequence (MS) of star formation (e.g. \citealt{Daddi2007}, \citealt{Elbaz2007}, \citealt{Rodighiero2011}, \citealt{Whitaker2012}, \citealt{Speagle2014}, \citealt{Schreiber2015}). 
First studies pointed out that SFR increases with stellar mass as a linear relation, with normalization varying according to the redshift (as well as the choice of initial mass function (IMF) and/or SFR indicators). Recent studies have found that this relation is linear for stellar masses up to $\sim$ 10$^{10}$ M$_{\odot}$ and actually flattens towards higher stellar masses (e.g. \citealt{Whitaker2014}, \citealt{Schreiber2015}). 

The most luminous AGN tend to reside in the most massive galaxy hosts, therefore the mutual dependence of AGN strength and SFR on stellar mass could lead to the correlation observed for AGN luminosity and SFR (e.g. as demonstrated by \citealt{Stanley2017}, \citealt{Yang2017}). Several studies have used X-ray luminosity as AGN strength indicator, however as discussed in \cite{Hickox2014}, it traces the instantaneous AGN activity on timescale much shorter than the timescale for star formation (>100 Myr). [OIII] luminosity instead, which is produced in the narrow line region, traces the AGN activity on longer timescales, resulting in a stronger correlation between the AGN luminosity and the SFR, as found by \cite{Zhuang2020}.

Other studies explored the AGN activity comparing the host galaxies properties with that of star-forming galaxies. Some find that AGN host galaxies mainly lie above or on the MS of galaxies (e.g. \citealt{Silverman2009}, \citealt{Santini2012}; \citealt{Mullaney2012}), whereas others find that most AGN host galaxies are below the MS, 
suggesting that AGN activity might regulate star formation inside their host galaxies through feedback mechanism (e.g. \citealt{Bongiorno2012}; \citealt{Mullaney2015}, \citealt{Shimizu2015}). These discordant results can be due to different AGN selection techniques, and SFR indicators. 

 Measuring AGN and star formation activity in these systems is therefore crucial to determine whether these processes are causally linked or not. 

AGN can be identified at different wavelengths, 
in the X-rays, mid-IR, radio and optical bands. The central source ionizes the gas located at kiloparsec scale distances, showing characteristic emission-line intensity ratios discernible from those coming from normal star-forming regions. Therefore one simple and physical method for classifying AGN is to examine their emission line ratios. AGN can be classified into two classes depending on whether the central engine is viewed directly (Type I) or is obscured by a dusty torus (Type II) (e.g. \citealt{Antonucci1993}, \citealt{Urry2000}). The spectra of Type I AGN show broad permitted emission lines (full width at half maximum (FWHM) $\ge$ 2000 km/s), originating from the so-called broad line region (BLR); on the contrary those of Type II AGN show narrow permitted and forbidden lines. Type II AGN can be identified in spectroscopic surveys by using the ratio of specific emission lines such as [OIII] over H$\beta$ and [NII] or [SII] over H$\alpha$ up to z$\sim$0.5 and [OII] over H$\beta$ at z$\geq$ 0.5 up to z$\sim$1.
 In Type I AGN the optical continuum is dominated by non-thermal emission, making it a challenge to study the host galaxy properties. We have therefore focused our analysis on Type II AGN.

The SFR can be estimated from the spectral energy distribution (SED) of an AGN, however the energy output can affect the entire SED, contaminating the SFR indicators usually used for the star-forming galaxies. Broadband SED and infrared band are frequently used to calculate SFR for X-ray selected AGN (e.g. \citealt{Mullaney2012}, \citealt{Stemo2020}).  The AGN contribution to the infrared luminosity, if not taken into account, would overestimate the infrared-based SFR (\citealt{Zhuang2018}) as well as other SFR indicators (e.g. \citealt{Azadi2015},\citealt{Ho2005}).
  Optical spectral features can be used to measure the stellar properties of host galaxies, such as the 4000 \AA\ break, the strengths of the H$\delta$ absorption (e.g. \citealt{Kauffmann2003c}) and [OII] emission line (e.g. \cite{Ho2005}, \citealt{Zhuang2019}). Indeed these have been extensively used to measure the properties of statistical samples of AGN host galaxies (e.g. \citealt{Kauffmann2003a}; \citealt{Silverman2009}; \citealt{Ho2005}).

\cite{Kauffmann2003a} have analysed a large sample of Type II AGN galaxies selected from the Sloan Digital Sky Survey (SDSS) at low redshift 0.02 < z < 0.3, and showed that AGN are typically hosted by massive galaxies (>3$\times 10^10 M\odot$) with properties similar to ordinary early-type galaxies, with spectral signatures of young stellar populations (10$^8$-10$^9$ yr) in AGN exhibiting high [OIII] luminosity (L$_{[OIII]}$>10$^7$ L$\odot$). Analysing SDSS DR7 galaxies which also include Type II AGN and LINERS, \cite{Leslie2016} show that AGN activity plays an important role in quenching star formation in massive galaxies. Indeed they find that the SFR in these objects is below the expected value according to the MS. 
\cite{Ho2005} used [OII]$\lambda$3727 as a tracer of ongoing star formation in a statistical sample of AGN, finding that optically selected AGN host galaxies exhibit low level of SFR despite the abundant molecular gas revealed. This finding suggests that such systems are less efficient in forming stars with respect to galaxies with similar molecular content, possibly due to the activity of the central nucleus.

We extend such analysis at higher redshift and to stellar masses lower the than typical values probed by SDSS survey. Using the VIPERS (e.g. \citealt{Guzzo2014},\citealt{Garilli2014}, \citealt{Scodeggio2018}) and VVDS spectroscopic surveys (e.g. \citealt{LeFevre2013}), we aim at investigating whether the SFR of the host galaxies, relative to that expected at a given stellar mass and redshift for a normal star-forming galaxy, changes as a function of AGN power and galaxy stellar mass, in a statistical sample of Type II AGN galaxies at 0.5 $\le$ z $\le$ 0.9, selected on the basis of their optical emission lines (\citealt{Lamareille2010}). 

The paper is organized as follows. In Sec. \ref{sec:vipers} and \ref{sec:vvds} we summarize the VIPERS and VVDS survey properties; in Sec. \ref{sec:analysis} the spectroscopic analysis along with the Type II AGN sample selection, as well as the SED fitting analysis are presented; Sec. \ref{sec:results} discusses the properties of Type II AGN host galaxies in the SFR-stellar mass plane, with the discovery of two distinct populations, along with their spectral properties, and a tentative evidence for AGN feedback that quenches star formation. Sec. \ref{sec:summary} provides a summary of the paper.

Throughout this work, we assume a standard cosmological model with $\Omega_M$ =0.3, $\Omega_{\lambda}$ =0.7, and H$_0$ =70 km s$^{-1}$Mpc$^{-1}$.

\section{The sample}
The goal of the present paper is to select and study the properties of narrow emission line AGN at intermediate redshifts (0.5 < z <0.9). For this purpose, we have collected spectroscopic and photometric data from the VIPERS survey (\citealt{Guzzo2014}, \citealt{Garilli2014}, \citealt{Scodeggio2018}) and the VVDS survey (\citealt{LeFevre2005}). The resulting sample is indicated as the VIMOS sample throughout the paper. 

In the local universe, the ionization source, either AGN or star formation, can be identified by using the intensity ratios of emission lines such as [OIII]$\lambda$5007, H$\beta$, H$\alpha$, [NII]$\lambda$6583, [SII]$\lambda\lambda$6717,6731 through specific diagnostic diagrams (\citealt{Baldwin1981}, BPT), which are accessible using ground-based optical telescopes up to z$\le$0.5. However, at higher redshift the H$\alpha$,[NII] and [SII] lines are redshifted in the NIR range  and can no longer be used, therefore alternative diagrams have been proposed. It is possible to use the [OII] emission line doublet, which enters the optical spectra at z >= 0.5, and the optical diagnostic diagram originally proposed by \cite{Rola1997} and further improved by \cite{Lamareille2010}, based on the ratios [OIII]/H$\beta$ vs. [OII]/H$\beta$ (also known as blue diagram).

\subsection{VIPERS survey}\label{sec:vipers}

The VIPERS spectroscopic survey was designed to sample galaxies at redshift 0.5 $\simlt$  z $\simlt$  1.2, selected from the Canada-France-Hawaii Telescope Legacy Survey Wide (CFHTLS Wide) over the W1 and W4 fields (\citealt{Guzzo2014}, \citealt{Garilli2014}, \citealt{Scodeggio2018}).
The observations were carried out using ESO VIsible Multi-Object Spectrograph (VIMOS) on Unit 3 of the ESO Very Large Telescope (VLT), with the low resolution red grism (R $\sim$ 210) and a slit width of 1 arcsec, covering the spectral range 5500-9500 \AA\ with a dispersion of 7.14 \AA\ per pixel. To achieve useful spectral quality in a limited exposure time, a bright magnitude limit of i$\rm_{AB}$ \simlt 22.5 was adopted, while low redshift (z $<$ 0.5) galaxies were removed using  the color-color selection in the (r-i) vs. (u-g) plane. Full information regarding observations, data reduction and selection criteria are contained in \cite{Garilli2014}, \cite{Guzzo2014}.

In this work, we use the final VIPERS data from the Public Data Release 2 (PDR-2, \citealt{Scodeggio2018}) containing 91507 galaxies with a measured redshift. 
To collect a reliable sample of sources that host AGN, we adopted 
the so-called blue diagram (\citealt{Lamareille2010}). 
From the PDR-2 VIPERS catalogue, we selected sources with highly reliable [OIII]$\lambda$5007,[OII]$\lambda$3726 and H$\beta$4861 ([OIII], [OII] and H$\beta$ hereafter) line measurements, i.e. 
satisfying the following constraints: the distance between the expected position and the Gaussian peak must be within 7 \AA\ ($\sim$ 1 pixel), the FWHM of the line must be between 7 and 22 \AA\ (from 1 to 3 pixels), the Gaussian amplitude and the observed peak flux must differ by no more than 30\% and the equivalent width (EW) must be detected at 3.5$\sigma$ or flux  at $\simgt$ 8$\sigma$. The final VIPERS sample consists of 7125 galaxies.

\subsection{VVDS survey}\label{sec:vvds}
We complemented the VIPERS data with the VIMOS VLT Deep Survey (VVDS). This survey was designed to study the evolution of galaxies, large scale structures and AGN in the redshift range 0 $<$ z $<$ 6.7 using the VIMOS spectrograph with the same instrument configuration as in VIPERS. 
VVDS is the result of a combination of magnitude-limited surveys as WIDE, which covers three fields (1003+01, 1400+05, 2217+00) down to I$\rm_{AB}$=22.5, DEEP, targeting the 0226-04 and ECDFS fields down to I$\rm_{AB}$=24 and Ultra-Deep (0226-04 field) in the magnitude range 23 < I$\rm_{AB}$ < 24.75. 
We have used the final VVDS dataset (\citealt{LeFevre2013}), collecting sources with reliable redshift estimates (with redshift flag z$\rm_{flag}=2,3,4$ corresponding to 75 up to 100 \% probability that the redshift is correct). Furthermore, we focused on targets with [OIII] and H$\beta$ fluxes detected at $\ge$ 5 and 2$\sigma$, respectively, and FWHM > 7 \AA\ (1 pixel), ensuring us to collect a clean sample as confirmed by visual inspection. The final VVDS sample consists of 1663 galaxies. 

The requirement to have both [OII] and [OIII] in the observed spectrum limits our sample to the redshift range 0.5<z<0.9.
Applying the above selections, our starting VIMOS sample comprises 8788 galaxies.

\section{Analysis}\label{sec:analysis}

\subsection{Spectroscopic analysis}
In order to obtain a uniform and precise measurement of line fluxes and equivalent widths, we subtracted the stellar continuum with absorption lines from the galaxy spectra, using the penalized pixel fitting public code (pPXF \citealt{Cappellari2012}). Specifically, the spectra are fitted with a linear combination of stellar spectra templates from the MILES library (\citealt{Vazdekis2010}, library included in the software package), which contains single stellar population synthesis models, covering the full range of the optical spectrum with a resolution of FWHM=2.54 \AA.   
We convolved the templates spectra with a Gaussian in order to match the spectral resolution of the VIMOS observed galaxy spectra, which have lower resolution. 
We included low order multiplicative polynomials to adjust the continuum shape of the templates to the observed spectrum. In the fitting procedure, the spectra are shifted to restframe and strong emission features are masked out. The pPXF best-fit model spectrum is chosen through $\chi^2$  minimization. Objects with best-fit results associated with reduced $\chi^2$ larger than 1 sigma of the $\chi^2$-distribution were discarded ($\sim$19\%). 

The residual spectrum obtained by subtracting the best-fit stellar model from the observed spectrum of each target is then used to characterize the emission line features. As mentioned in Sec. \ref{sec:vipers}, we used [OII], [OIII], H$\beta$ to identify AGN Type II using optical diagnostic tools.

For each spectrum, we have adopted as systemic redshift the one estimated from the [OIII]$\lambda$5007 emission line, and
we performed a fit \footnote{
The spectral analysis was performed with the python routine {\tt{scipy.optimize.curve\_fit}}.} of stellar-subtracted spectra shifted to the rest-frame.
 We separately fit two spectral regions, focusing on the [OIII]-H$\beta$ and the [OII] doublet lines. We adopted a linear function to model possible continuum residuals, while Gaussian components were used to reproduce the emission lines.

We fixed the wavelength separation and broadening between the [OIII]$\lambda$5007 \AA\ and [OIII]$\lambda$4959 \AA\ , H$\beta$ lines. The flux intensities of the [OIII] doublet is set to 1:3, according to their atomic parameters (\citealt{Osterbrock2006}).
The [OII] emission line doublet is unresolved in our spectra and we fit its profile using a single Gaussian model, with three free parameters (normalization, centroid and sigma).

From the best-fit models, we derive as spectral line parameters the flux and the EW.

We note that 73\% of the VVDS sample presented here was already analysed by \cite{Lamareille2009}.  We checked for consistency between our line measurements and those presented in \cite{Lamareille2009}, and found a fair 
agreement, with median absolute differences between EW being 0.8 \AA, 2.2 \AA\ and 0.3 \AA, for [OIII], [OII] and H$\beta$ respectively. 

\subsection{Identification of Type II AGN}\label{sec:selection}
\begin{figure}[]
 \includegraphics[width=1.1\columnwidth]{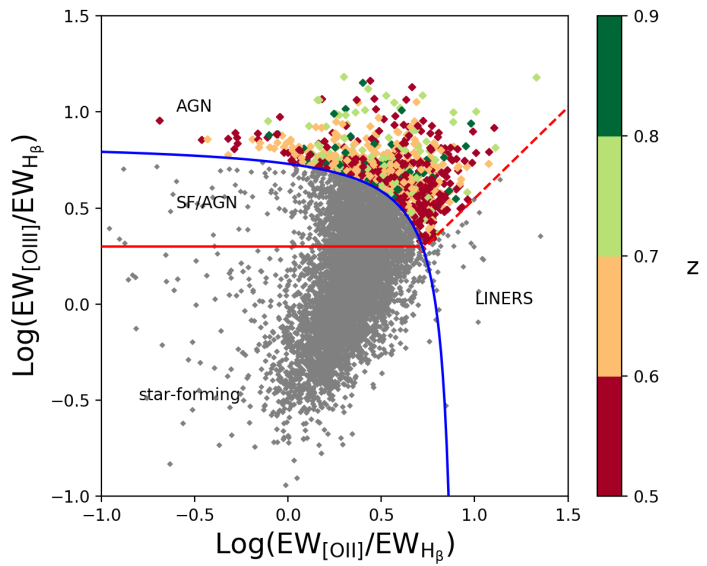}

 \caption{\small{Blue diagram (\citealt{Lamareille2010}) for the VIMOS sample. Diamonds represent Type II AGN color-coded according to their redshift. VIMOS galaxies with reliable emission line measurements are marked with grey dots. The blue curve shows the separations between starforming galaxies and AGN (Eq. \ref{eq:AGN}), the red dashed line between AGN and LINERs (Eq. \ref{eq:LINER}), the red solid line between starforming galaxies and SF/AGN (Eq. \ref{eq:composite}).}}\label{fig:BPT}
\end{figure}

The demarcation proposed by \cite{Lamareille2010} to separate star-forming galaxies (SFG) and Type II AGN (shown as blue curve in Fig. \ref{fig:BPT}) is the following:

\begin{equation}
\rm log\frac{[OIII]}{H\beta}=\frac{0.11}{log\frac{[OII]}{H\beta}-0.92}+0.85. \label{eq:AGN}
\end{equation}

The boundary used to distinguish between the Type II AGN and the LINERS regions (shown as red dashed line in Fig. \ref{fig:BPT}) is:

\begin{equation}
\rm log\frac{[OIII]}{H\beta}= 0.95 \times log\frac{[OII]}{H\beta}-0.4, \label{eq:LINER}
\end{equation}

and the region where SFGs are mixed with AGN (shown as red solid line in Fig. \ref{fig:BPT}) is given by:
\begin{equation}
\rm log\frac{[OIII]}{H\beta} >0.3. \label{eq:composite}
\end{equation}

Given the wavelength distance between [OIII] and H$\beta$ on one side, and [OII] doublet lines on the other, these line ratios are sensitive to reddening. \cite{Lamareille2010} have demonstrated that the use of equivalent widths instead of the line fluxes minimizes this problem, even if it does not eliminate it completely. We thus use EW ratios instead of flux ratios to minimize the effect of reddening.


In Fig. \ref{fig:BPT} the blue diagram as found for the VIMOS sample is shown. Our selection of AGN Type II includes 812 objects. 

\subsection{Ancillary data}
In order to carry out our study, we need to estimate galaxy properties such as stellar masses and star formation rates for the VIMOS AGN sample.  The study of broad-band spectral energy distribution (SED) is the most commonly method adopted to derive galaxy properties.
To estimate the selected AGN stellar mass we collect all available photometric data and fit them with galaxy$+$AGN templates.
\subsubsection{VIPERS photometry}\label{sec:photometry}


The VIPERS sample has been selected from the W1 and W4 fields of the CFHTLS, which provides magnitudes in the {\it{u$^*$,g,r,i,z}} photometric bands down to i < 22.5, 
corrected for Galaxy extinction derived from the Schlegel dust maps (\citealt{Guzzo2014}, \citealt{Moutard2016}). The following additional photometric data are available for a subset of sources:
\begin{itemize}
    \item  NIR observations are available for 98\% of the AGN sample in the Ks band in the W1 and W4 fields and in the K$\rm_{video}$ band in the W1 field (\citealt{Moutard2016}). 
    
    \item the VIPERS survey is also covered by GALEX observations in the FUV and NUV bands for 17\% of the AGN sample (\citealt{Moutard2016}).
    
    \item MIR photometry is available for 13\%  of the AGN targets with Spitzer, from the Spitzer WIDE-area Infrared Extragalactic survey observations in the XMM-LSS field (\citealt{Lonsdale2004})
    
    \item Photometric information in the WISE all-sky passbands is also available for 18\% of the AGN targets (\citealt{Wright2010}, VIPERS team). 
\end{itemize}






\subsubsection{VVDS photometry}\label{sec:photometry_VVDS}

All the VVDS fields have been observed in B, V, R, I filters as part of the VIRMOS Deep Imaging Survey (\citealt{McCracken2003}; \citealt{LeFevre2004}) with the CFH12K imager at CFHT. The following additional photometric data are available:

\begin{itemize}
    \item u*, g', r', i', z' photometry from the Canada-France-Hawaii Telescope Legacy Survey (CFHTLS, \citealt{Cuillandre2012}) is available for 43\% of the AGN sample and NIR photometric information from WIRcam InfraRed Deep Survey in the J, H, and K bands (\citealt{Bielby2012}) and from UKIDSS in the J and K filters (\citealt{Lawrence2007}) for 38\% and 44\% of the AGN sample, respectively.
    \item  FUV and NUV photometry with the GALEX satellite (\citealt{Arnouts2005}) is available for 2\% of the AGN sample and MIR data with Spitzer (SWIRE survey, \citealt{Lonsdale2003}) for 3\% of the sample.
    \item Imaging with Advanced Camera for Surveys Field Channel instrument on board the Hubble Space telescope is available for 4\% of the AGN sample, in four bands $B,V,I$ and $Z$ (\citealt{LeFevre2004b}).

\end{itemize}



\subsection{SED fitting analysis}\label{sec:SED}
We have derived  stellar masses and SFRs through SED fitting  of the available multiwavelength photometry (see Sec. \ref{sec:photometry}) using the 
Code Investigating GALaxy Emission (CIGALE; \citealt{Noll2009}, \citealt{Boquien2019}, version 2020.0).\footnote{CIGALE can also handle upper limits.} 

CIGALE provides a multi-component SED fit that includes multiple stellar components (old, and young), dust and interstellar medium (ISM) radiation, and AGN emission. The different components are linked to balance the absorbed radiation at UV-optical wavelengths with that re-emitted in the FIR.

The components used for the fitting procedure are (i) the stellar emission which dominates the wavelength range 0.3 - 5 $\mu$m, (ii) the emission by the cold dust, which is heated by the star formation and dominates the FIR and (iii) the AGN emission, coming from the accretion disk, peaking at UV-optical wavelengths and reprocessed by the dusty torus in the MIR. In the fitting procedure, we have fixed  the redshift at the value derived from the [OIII] emission line (see Sec. \ref{sec:analysis}).
The models adopted for the SED fitting are the following:
\begin{itemize}

\item For the stellar models we adopted a delayed star formation history (SFH), $\tau$-model, with varying e-folding time and main stellar population ages, defined as:
\begin{equation}
\rm SFR(t) \propto\ t \times exp(-t/\tau)
\end{equation}

where $\tau$ is the e-folding time of the star formation. 

\item The SFH is convolved with the stellar library of \cite{Bruzual2003} assuming the Chabrier initial mass function (IMF). The  metallicity is fixed to solar value, 0.02. We set the separation between the young and old stellar populations to 10 Myr. 

\item Dust extinction is modelled by assuming the \cite{Calzetti2000} law. We use the color-excess E(B-V)$_*$ in the range of values shown in Table \ref{tab:parameters}. We assume that old stars have a lower extinction compared to the young stellar populations by a factor of 0.44 (\citealt{Calzetti2000}). 

\item We adopted the \cite{Dale2014} templates to model the reprocessed emission in the IR from the dust heated by stellar radiation. These templates also include the contribution from the dust heated by AGN. We used them without the AGN contribution, which is set equal to 0, since it is defined separately with the \cite{Fritz2006} templates. The models represent emission from dust which is exposed to different ranges of radiation field intensity and the templates are combined to model the total dust emission with the relative contribution given by a power law distribution dM$\rm_{dust}$ $\propto$ U$^{-\alpha}$dU, with M$\rm_{dust}$ the dust mass heated by a radiation field and U the radiation field intensity. The free parameter $\alpha$ slope was allowed to vary in the range listed in Table \ref{tab:parameters}.

\item To parametrize the AGN emission component, we used the models from \cite{Fritz2006}, which assume isotropic emission from the central AGN and emission from the dusty torus. The law describing the dust density within the torus is variable along the radial and polar coordinates:
\begin{equation}
\rm \rho(r,\theta) = \alpha r^{\beta} e^{-\gamma|cos(\theta)|}
\end{equation}

with $\alpha$ proportional to the equatorial optical depth at 9.7 $\mu$m ($\tau_{9.7}$), $\beta$ and $\gamma$ are related to the radial and angular coordinates respectively. We fixed the parameters $\beta$, $\gamma$ and $\theta$ to parametrize the dust distribution within the torus, according to the values reported in Table \ref{tab:parameters}. The geometry of the torus is described by using the ratio between the outer and inner radii of the torus, R$\rm_{max}$/R$\rm_{min}$ and the opening angle of the torus, $\theta$. We choose typical values as found in \cite{Fritz2006} and by fixing these parameters we avoid degeneracies in the templates. It is possible to provide a range of inclination angle between the line of sight of the observer and the torus equatorial plane, $\psi$, with values ranging from 0 for Type II up to 90 for Type I AGN.
Another important parameter is the fractional contribution of the AGN emission to the total IR luminosity, frac$\rm_{AGN}$. We set a wide range of values to account for the possibility that the AGN contribution to the IR luminosity is very low, 5\%, up to 95\% of the total contribution.

\end{itemize}

\begin{figure}[]
 \includegraphics[width=1.1\columnwidth]{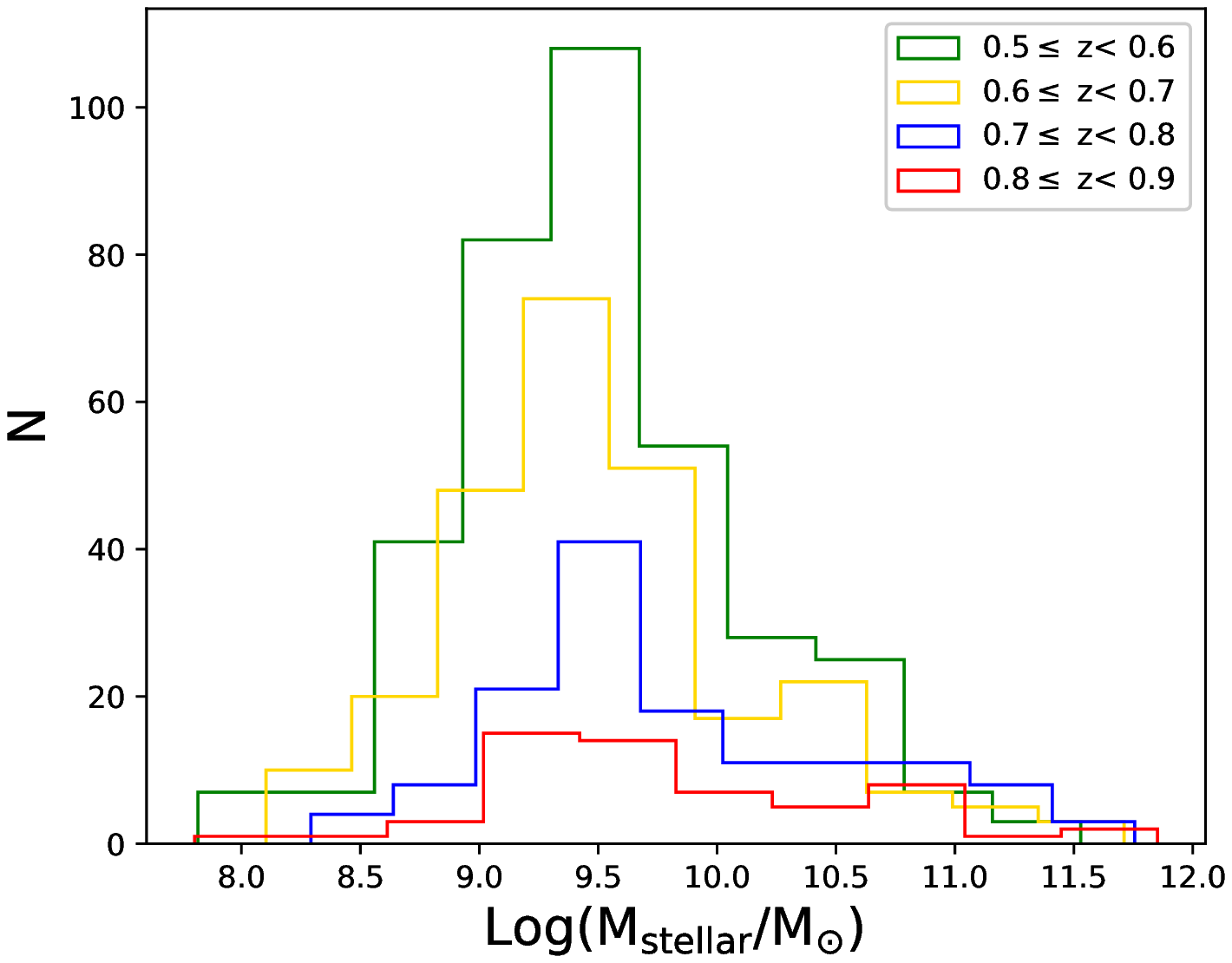}
 \caption{Stellar mass distribution of the Type II AGN host galaxies for the VIMOS sample in different redshift ranges}\label{fig:stellar_mass_redshift}
\end{figure}

The photometric data are fitted with the models and the physical properties are then estimated through the analysis of the likelihood distribution. In Fig. \ref{fig:stellar_mass_redshift} the distribution of stellar masses for AGN Type II is shown for different redshift bins. We probed stellar masses, Log (M$\rm_{stellar}/ M_{\odot}$), in the range $\sim$8-12, with a median (mean) value of 9.5 (10.2). 

\begin{table*}
\setlength{\tabcolsep}{0.05pt}

        \centering
        \begin{threeparttable}
        \caption{CIGALE parameters used for the SED fitting.}\label{tab:parameters}
                \begin{tabular}{ccc}
                        \hline
                        \hline
                        Parameter & Description & Value \\ 
                        \hline
                        & Star Formation History - Delayed Model & \\
                        \\
                               Age &     Age of the main stellar population     &   500, 1000, 3000, 4000,5000\\
                               &&  5500, 6000, 7000, 8000, 9000 Myr \\  
                $\tau$  & e-folding time of the main stellar population & 0.5, 1.0, 3.0, 5.0, 10.0 Gyr  \\
\hline
 &\cite{Bruzual2003} Stellar Emission Model&\\
 \\
	   IMF &      Initial mass function     & Chabrier \\  
 Z &  Metallicity  & 0.02  \\
   Separation age &   Separation between the young and the old stellar population & 10 Myr  \\
                            \hline

              & \cite{Calzetti2000} and \cite{Leitherer2002} Dust attenuation model &\\
              \\
              E(B-V) & Colour excess of the young stellar continuum light &  0.05,0.1,0.3,0.5,0.7,0.9,1.1,1.3  \\
                        UV bump & Amplitude of the UV bump & 0.0\\
                        Slope & Slope delta of the power law attenuation curve & 0.0 \\
                        reduction factor  & Reduction factor for the color excess of & 0.44\\
                        & the old population compared  to the young one  &\\
                        \hline
& Nebular Emission model &\\
\\
              U & Ionization parameter& 10$^{-2}$ \\
              f$\rm_{esc}$& Escape fraction of Lyman continuum photons & 0\%\\
              f$\rm_{dust}$& Absorption fraction of Lyman continuum photons & 10\%   \\

              \hline
              & \cite{Dale2014} Dust Module &\\
              \\
               $\alpha$ & Slope of the power law combining the contribution of different dust templates & 0.5, 1.0, 1.5, 2.0, 2.5, 3.0  \\
              \hline
             & \cite{Fritz2006} AGN Module & \\
             \\
              R$\rm_{max}$/R$\rm_{min}$ & Ratio of the maximum to minimum radii of the dust torus &60  \\
              $\tau_{9.7}$ & Optical depth at 9.7 $\mu$m &1.0 \\
 $\beta$  &               Slope of the radial coordinate & -0.5 \\

        $\gamma$ & Exponent of the angular coordinate &0.0 \\
       $\Phi$ & Full opening angle of the dust torus & 100 deg  \\
        $\psi$ & Angle between equatorial axis and line of sight & 0.001, 10.1, 20.1, 30.1,50.1,70.1 \\
       f$\rm_{AGN}$  & AGN fraction & 0.05,0.1,0.15,0.2,0.25,0.3,0.35,0.4,\\
       &&0.45,0.5,0.55,0.6,0.65,0.7,0.75,0.8, \\
       &&0.85 ,0.9,0.95 \\
                        \hline
                        
                \end{tabular} 
                                       \end{threeparttable}
        
\end{table*} 

\subsubsection{Mock analysis}

\begin{figure}[]
 \includegraphics[width=1.1\columnwidth]{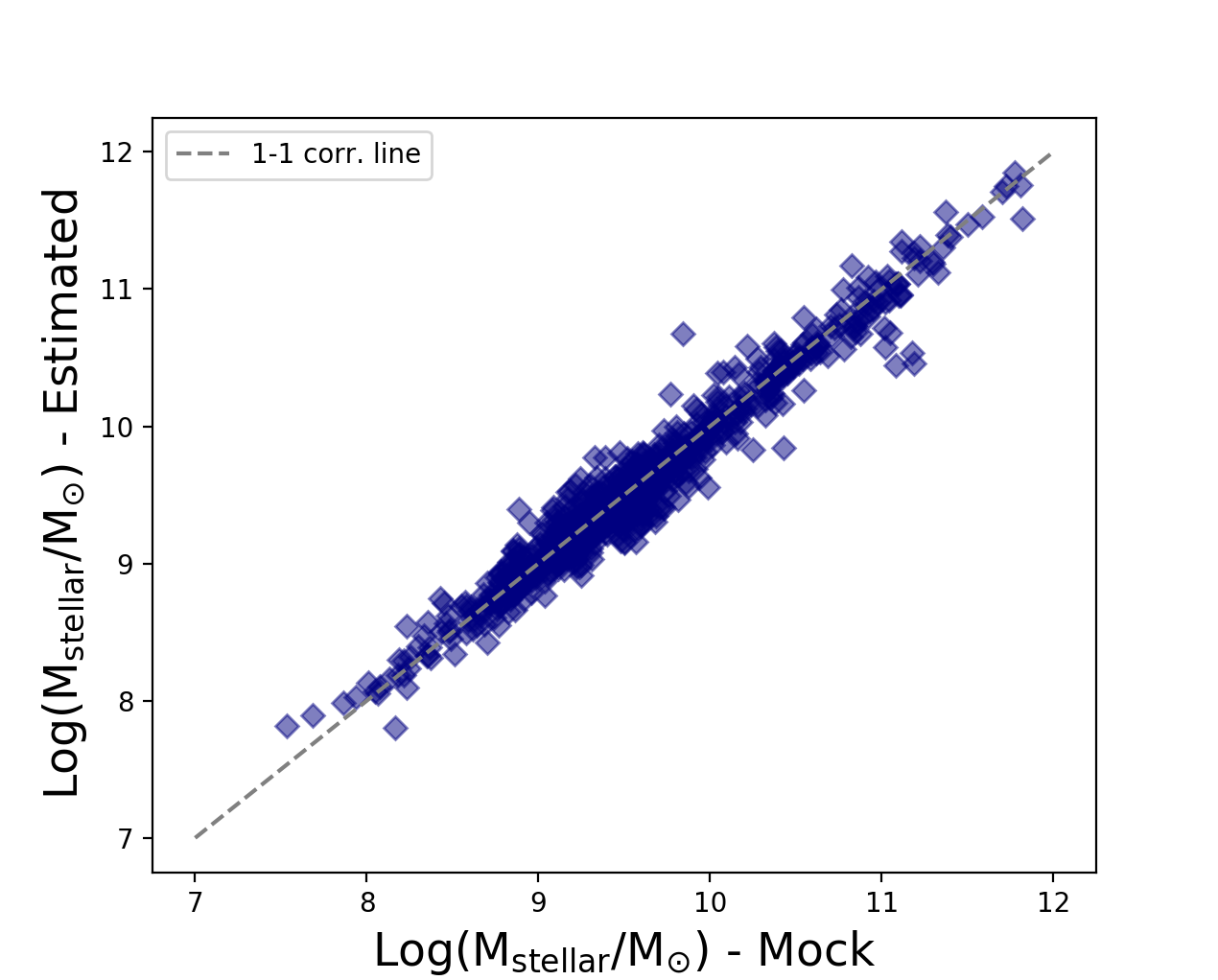}

 \caption{Comparison between the true value of the stellar mass as derived from the mock analysis and the value estimated by SED fitting. The grey dashed line indicates the 1:1 relation between the parameters.}\label{fig:mstar_mock}
\end{figure}

The reliability of the computed host galaxies stellar mass values from the SED fitting analysis can be assessed through the analysis of a mock catalogue. The basic idea is to compare the stellar masses of the mock catalogue (true values), which are known exactly, to the values estimated from the analysis of the likelihood distribution. We use an option included in CIGALE to build a mock catalogue, based on the best-fit model for each object, as derived in Sec \ref{sec:SED}. A detailed description of the mock analysis can be found in \cite{Giovannoli2011}. Briefly, the best-fit SED model of each object is modified by adding a randomly Gaussian-distributed error to each flux measured in the photometric bands of the dataset, with the same standard deviation as the observed uncertainty in each band. The mock catalogue is then analysed in the exact same way as the real observations. Fig. \ref{fig:mstar_mock} shows the comparison between the stellar masses derived from the mock analysis and the values estimated for the real sample of Type II AGN. The estimated and true values are well correlated, indicating that the stellar mass parameter can be consistently constrained, with a Pearson's correlation coefficient for linear regression of $\sim$0.98.

\subsection{SFR}


From the SED decomposition it would also be possible to derive the SFR, however a lack of FIR coverage prevents us to retrieve a reliable estimate of the SFR from the SED (see \citealt{Ciesla2015}).
An alternative SFR indicator is the [OII]$\lambda$3726+3728 doublet line.
The [OII] emission line is commonly used to measure SFR in star forming galaxies (e.g. \citealt{Kennicutt1998}, \citealt{Hopkins2003}, \citealt{Kewley2004}), despite it suffers from dust-extinction, as it is known to be strongly excited  by star formation. 
For galaxies with an active nucleus, lines of low ionization potential such as [OII] could be excited by both SF and AGN activity, however it has been observed that the [OII] is mainly produced by star-formation (e.g. \citealt{Ho2005}, \citealt{Zhuang2019}).

\begin{figure}[]
 \includegraphics[width=1.1\columnwidth]{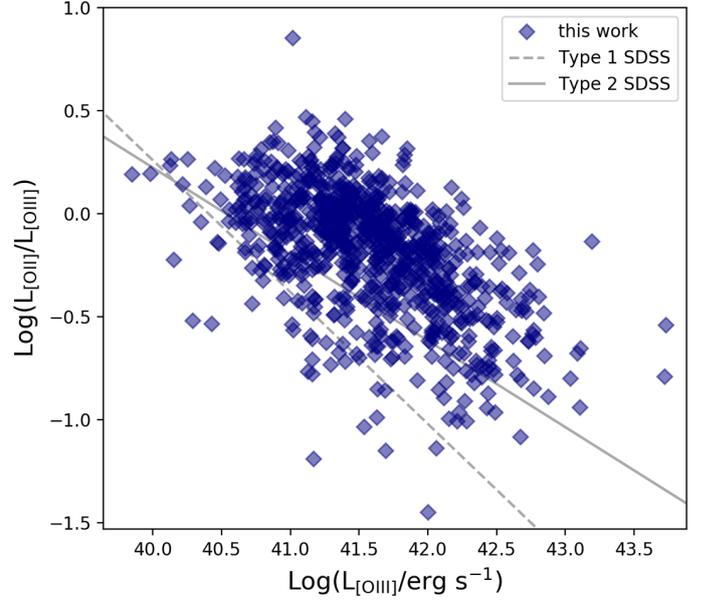}

 \caption{\small{[OII]/[OIII] luminosity ratio as a function of [OIII] luminosity for the VIMOS Type II AGN. Dashed and solid lines represent the best-fit relation for SDSS Type I and Type II AGN at z<0.3, respectively. }}\label{fig:LOII_LOIII}
\end{figure}

As discussed in \cite{Silverman2009}, the [OII]/[OIII] flux ratio decreases at increasing [OIII] luminosities and the slope that describes this relation is flatter for Type II AGN than for Type I AGN (see Fig. \ref{fig:LOII_LOIII}). This difference is explained by an additional contribution to the [OII] flux in type II AGN due to on-going star formation. Previously \cite{Kim2006} explained the enhanced [OII]/[OIII] ratios for Type II Quasars from \cite{Zakamska2003} (median value of [OII]/[OIII]= $-$0.12) as due to a more prevalent star formation in Type II AGN. 
In Fig.  \ref{fig:LOII_LOIII}  we show the [OII]/[OIII] luminosity ratio as a function of [OIII] luminosity for the VIMOS sample. The median value of the [OII]/[OIII] ratio is $-$0.14, consistent with the Zakamska et al. sample. We note that the line luminosities are not corrected for extinction, therefore the line ratios can be considered as lower limits. 
We also plot the best-fit linear relation for SDSS Type I (dashed line) and Type II (solid line) sources as reported in \cite{Silverman2009}.   
Type II AGN in the VIMOS sample exhibit a similar slope to the SDSS Type II AGN and have slightly enhanced [OII]/[OIII] ratios compared to the SDSS sample. This finding further justifies the use of this line as SFR indicator (e.g. \citealt{Silverman2009}, \citealt{Kim2006}).

Assuming that high ionization lines such as [OIII] are mainly powered by AGN activity (e.g. \citealt{Kauffmann2003a}), we adopted this line to remove the AGN contribution from the [OII] line.
 Recently \cite{Zhuang2019} found a fairly constant [OII]/[OIII]$\sim$0.10 for the Type II AGN contribution according to a set of photoionization models, we therefore subtracted 10\% of the [OIII] luminosity from the [OII] line. We then derived the SFR by using the calibration from  \cite{Kewley2004} (rescaled by a factor of 1.7 to account for the different IMF used): 
 
\begin{equation}
\rm SFR_{[OII]} = 6.58 \pm 1.65 \times\ 10^{-42} (L_{[OII]} - 0.109L_{[OIII]}) \\    (M_{\odot} yr^{-1})\label{eq:SFR}
\end{equation}

where L$\rm_{[OII]}$ and L$\rm_{[OIII]}$ are in units of erg/s.
We probed SFR in the range 0.01-38 M$\odot$/yr, with a median(mean) value of 0.8 (1.3) M$\odot$/yr.

\section{Results}\label{sec:results}

\subsection{SFR-stellar mass plane}\label{sec:sfr_mass}

\begin{figure}[]
 \includegraphics[width=1.1\columnwidth]{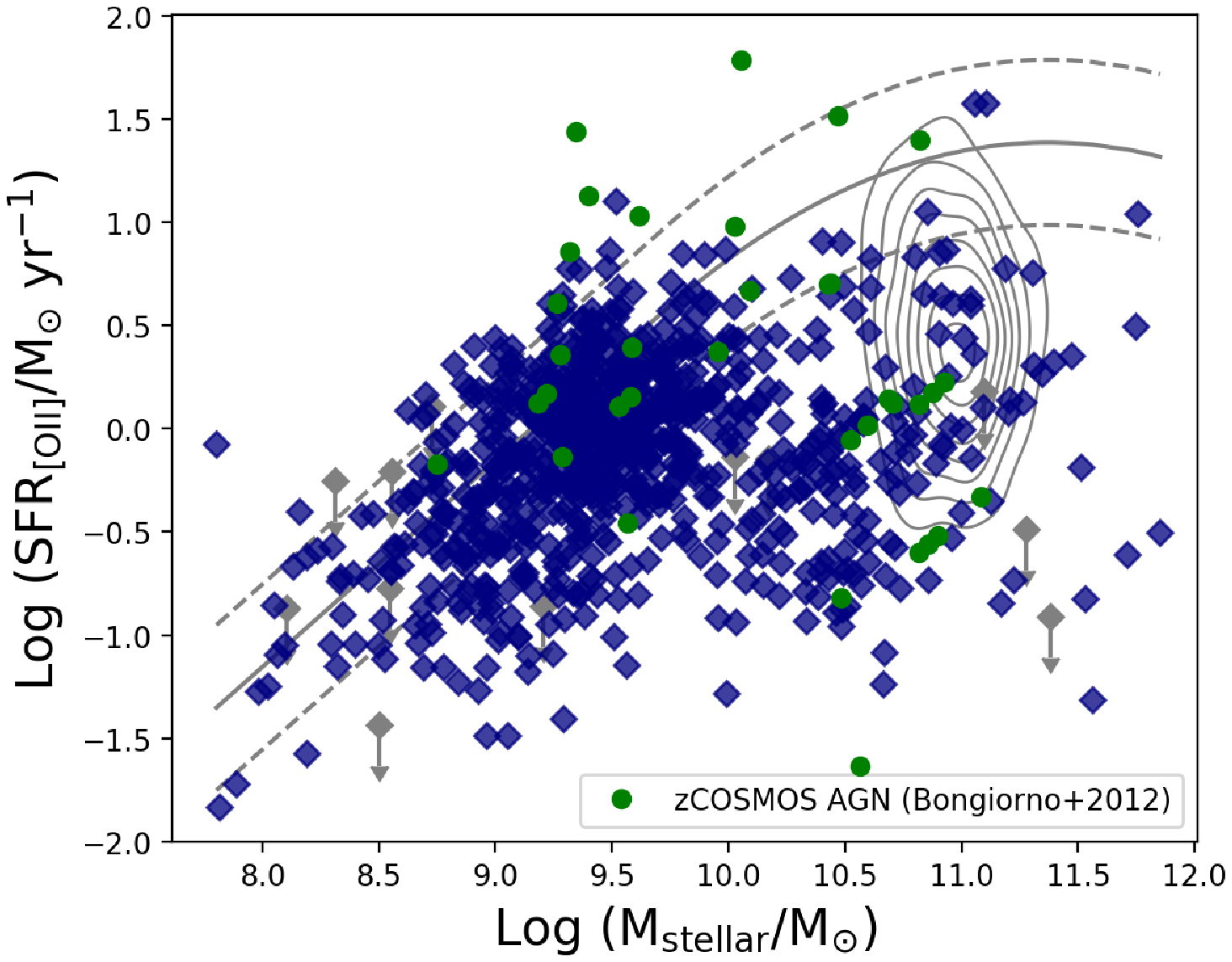}
\caption{Star formation rate (SFR) as a function of stellar mass (M$\rm_{stellar}$) for the VIMOS sample (blue diamonds). The solid line represents the star-forming main sequence as found by \cite{Schreiber2015} and dashed lines mark the scatter of 0.4 dex. Upper limits are shown as grey diamonds. Optically selected Type II AGN from zCOSMOS survey as found by \cite{Bongiorno2012} and from BOSS DR12 are shown as green circles and grey contours, respectively.}\label{fig:SFR_Mstar}
\end{figure}

 In Fig. \ref{fig:SFR_Mstar}, we show the SFR-stellar mass relation for the Type II AGN host galaxies of the VIMOS sample. We indicate the star-forming MS relation at z=0.7, the mean redshift of our sample, from \cite{Schreiber2015} (solid curve) along with the scatter (0.4 dex, dashed lines).  We rescaled both the SFR and stellar masses of \cite{Schreiber2015} by a factor of 1.7 to account for the different IMF used (Salpeter vs. Chabrier). 
 The bulk of the VIMOS sample populates the MS region, with a fraction of AGN host galaxies on and off the MS. 
 
 At high stellar masses (> 10$^{10}$ M$_{\odot}$) almost all sources are below the MS (see Fig. \ref{fig:SFR_Mstar}). 
Overall, Type II host galaxies show a broader distribution of SFR than star-forming MS galaxies, consistent with previous studies (e.g. \citealt{Mullaney2015}).

As discussed in \cite{Bongiorno2012}, optically selected Type II sources from the zCOSMOS-bright survey show properties similar to the VIMOS sample. This is not surprising since both surveys cover similar volumes and depths. We add this sample in Fig. \ref{fig:SFR_Mstar} as green circles. 
These sources are selected through the blue diagram in the redshift range 0.50<z<0.92, with stellar masses and SFR derived through SED fitting analysis. In terms of stellar masses and SFR, they span the same range as the VIMOS Type II AGN galaxies, and also for these sources the MS locus at high stellar masses remains under-populated for Type II AGN host galaxies, with respect to what is found for star forming galaxies, suggesting that the distribution could be different with respect to non-AGN galaxies. 

We investigate whether a fraction of Type II AGN can be missed by the adopted selection criteria.
One possibility is that the "missing" fraction could reside in the composite locus of SF-Type II AGN. Indeed, a high level of star formation can produce an enhancement of the H$\beta$ flux, moving a Type II AGN down to the composite locus in the blue diagram plane. We therefore investigated the host-galaxies properties of the composite sources, as defined by the blue-diagram, in the VIPERS sample. 
We collected their SED-based stellar masses and measured the SFR using the [OII]$\lambda$3727 emission line.  We found that the composite galaxies actually exhibit stellar masses <10$^{10}$ M$\odot$ and SFR slightly enhanced with respect to those observed for Type II AGN in a similar mass range. This indicates that the missing fraction of Type II AGN is not classified as composite. 





We also compare the VIMOS sample with the DR12 BOSS sample of Type II AGN (\citealt{Thomas2013}). Specifically we used the galaxy properties (i.e. emission line measurements, BPT classification and stellar masses) derived by the Portsmouth Group for the BOSS DR12 (\citealt{Thomas2013}). They applied the blue diagram criterion to select a Type II AGN sample at redshift 0.5 < z < 0.9. We restricted our analysis to those galaxies with [OIII], [OII] and H$\beta$ flux detection over 2 sigma (as defined by the amplitude-over-noise ratio parameter). For consistency with the VIMOS sample, we measured SFRs of the Type II AGN host galaxies from the BOSS DR12 using the [OII]$\lambda$3727+3729 emission line fluxes, subtracting off the AGN contribution by using the [OIII]$\lambda$5007 line flux (see Eq. \ref{eq:SFR}). 
Stellar masses are calculated by the Portsmouth team from the best-fit SED (\citealt{Maraston2013}). They used two types of templates to derive stellar masses, i.e. passively evolving  and star-forming models, based on the galaxy types expected according to the BOSS color cut. As presented in \cite{Thomas2013}, BOSS Type II AGN preferentially show a g-r color (strongly dependent on the star formation history of galaxies) in between the one observed for luminous red and star forming galaxies. Here we used stellar masses from the star-forming model, noting that stellar masses could be underestimated.

A dust-extinction effect could still play a role. In case of highly-star forming galaxies (and high stellar mass) with an AGN, H$\beta$ emission line could remain undetected due to attenuation by dust and hence the galaxy would be excluded from our sample selection. This would result in a missing fraction of galaxies with high stellar mass. We therefore compare our sample with the Type II AGN from the BOSS survey. The bulk of Type II AGN is preferentially found in host galaxies with stellar mass > 10$^{10}$ M$\odot$, due to the BOSS selection color cut favouring the most massive galaxies and 17\% of Type II AGN from the BOSS survey occupies the locus of MS and starburst. Despite the 
slightly 
enhanced statistics than that probed by the VIMOS targets, we can conclude they are overall consistent, considering the small area covered by VIPERS and VVDS surveys (24 deg$^2$ for VIPERS, 8.7 deg$^2$, 0.74 deg$^2$, and 512 arcmin$^2$ for VVDS-Wide, Deep and Ultra Deep respectively) with respect to the BOSS survey ($\sim$10000 deg$^2$).

\subsection{Correlation between SFR offset from the Main Sequence and AGN luminosity}\label{sec:populations}

AGN feedback could be responsible for the quenching of star formation in massive galaxies, with increasing AGN efficiency in driving outflows at increasing AGN luminosity (e.g. \citealt{Menci2008}, \citealt{Faucher2012}, \citealt{Hopkins2016}). 

\begin{figure*}
\begin{minipage}{.5\linewidth}
\centering
\includegraphics[scale=0.45]{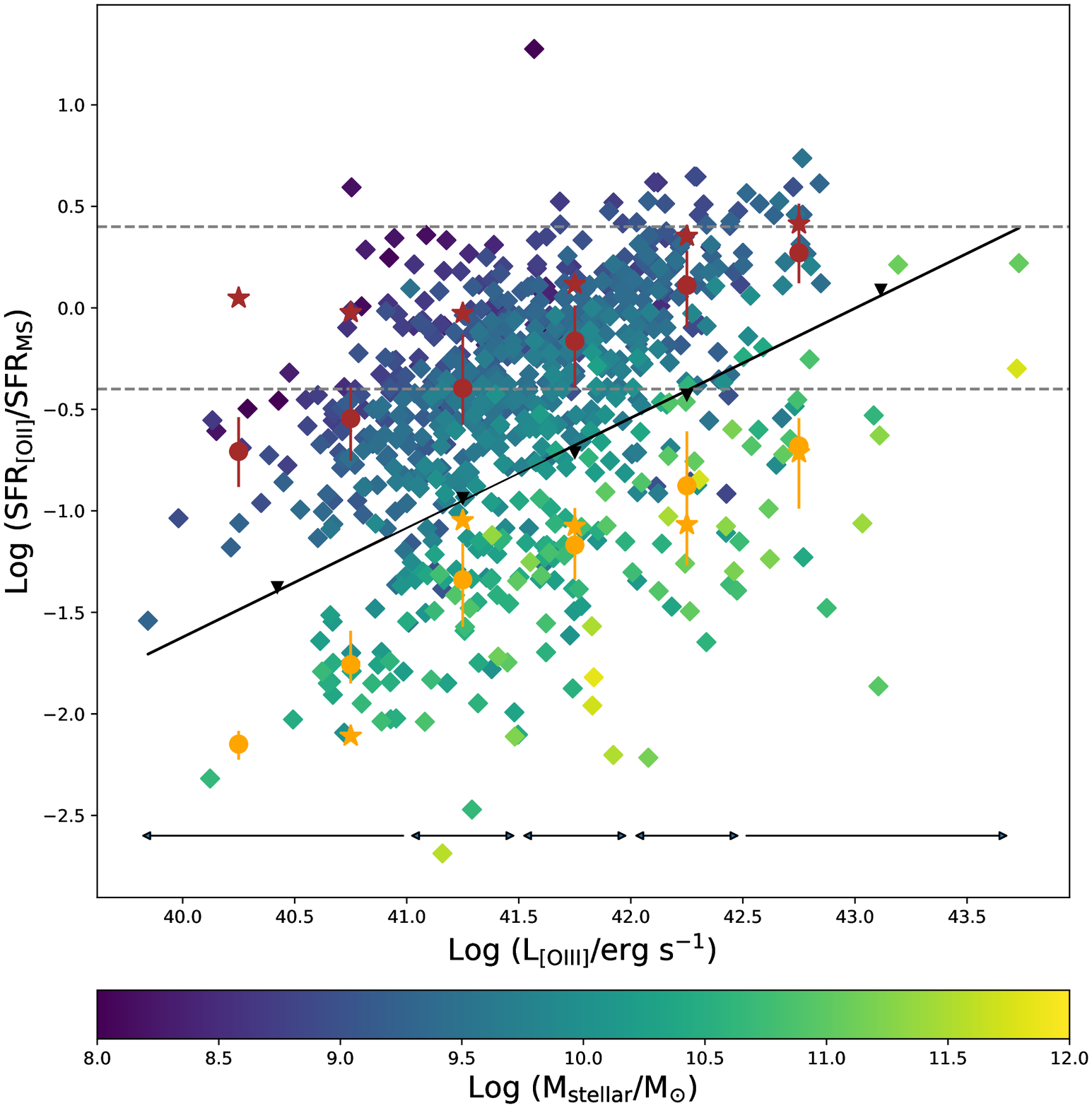}

\end{minipage}%
\begin{minipage}{.5\linewidth}
\centering
\includegraphics[scale=0.38]{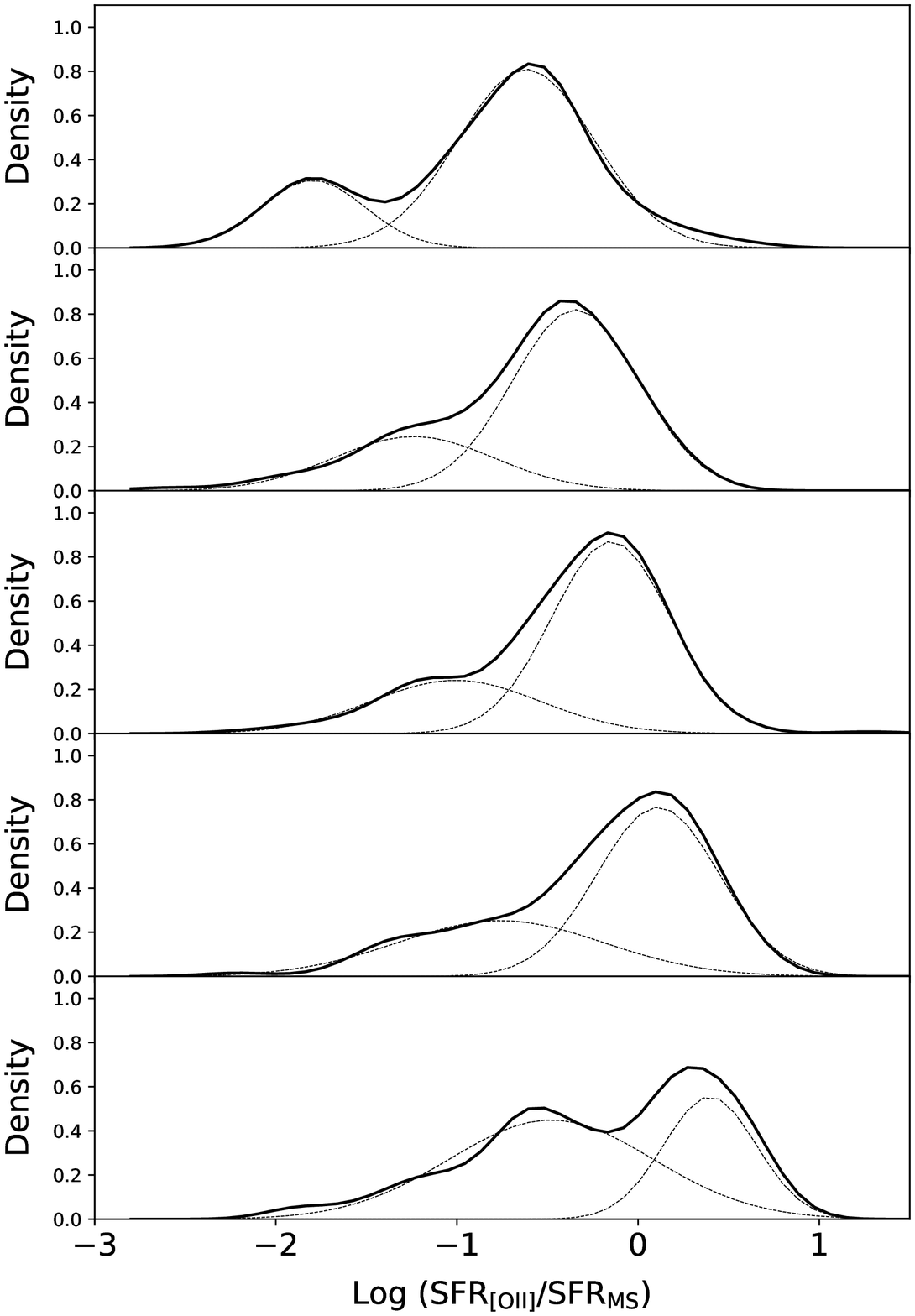}

\end{minipage}
\caption{(left panel) Distance from the MS (parameterised as SFR$\rm_{[OII]}/SFR_{MS}$) as a function of [OIII] luminosity (proxy of AGN power) for the VIMOS sample (diamonds) color-coded according to the stellar mass. 
Median values of [OII]-based and SED-based SFR offset in bins of [OIII] luminosities are represented as brown and orange circles and stars, below and above the line threshold, respectively (see the main text for details). The dashed lines delimit the locus of the MS ($\pm$0.4 dex). (right panel) Probability density function of the SFR offset in each luminosity bin (black solid lines), with superimposed the best-fit Gaussian components (black dashed lines) which reproduce the observed bimodality (see Sec. \ref{sec:populations}).}\label{fig:SFR_LOIII} 
\end{figure*}

To test this point, we examined the relationship between the distance from the MS and the AGN power. 
In Fig. \ref{fig:SFR_LOIII} we show the relative offset of the SFR from the MS relation of \cite{Schreiber2015} as a function of [OIII] luminosity, which can be considered as a proxy of AGN power.
VIMOS Type II AGN host galaxies show different properties in terms of star formation with a clear dependence on stellar mass (color-coded), forming two distinct groups of AGN on this diagram: at a fixed AGN power, Type II AGN host galaxies at M$\rm_{stellar}$< 10$^{10}$ M$_{\odot}$ show higher star formation activity than more massive galaxies. To define the boundary of this bimodality, we divided our targets in five subsamples with different AGN power (indicated in Fig. \ref{fig:SFR_LOIII}, left panel). 
In each luminosity bin, we used the Gaussian kernel density estimation (KDE) to estimate the probability density function of the SFR offset and derive the separation between the two subsamples (see right panel of Fig. \ref{fig:SFR_LOIII}). We proceeded as follows: i) we fit two Gaussians to reproduce the bimodality of KDE functions; ii) we derive the intersection point of the two best-fit Gaussians in each bin and iii) perform a linear regression on the intersection points. The best-fit line to these points, i.e. 0.54$\times$ Log (L$\rm_{[OIII]}$/\ergs) - 23.22, is shown in Fig. \ref{fig:SFR_LOIII}, left panel, as black solid line.

Above the boundary line, 64\% of the VIMOS subsample occupies the same region as the star-forming (i.e. log($\rm SFR_{[OII]}/SFR_{MS}$) within $\pm$0.4 dex) and starburst galaxies (i.e. log($\rm SFR_{[OII]}/SFR_{MS}$) > 0.4), with stellar mass mostly below 10$^{10}$ M$_{\odot}$, and the remaining 36\% have SFR below the bulk of the MS galaxies but above the quiescent locus (sub-MS, $-$1.3<log($\rm SFR_{[OII]}/SFR_{MS}$) < $-$0.4). Hereafter we will refer to this subsample as high-SF Type II AGN.
Instead below the line threshold the diagram is occupied by massive targets along the MS (3\%), 51\% in the sub-MS and 46\% occupy the quiescent regime (i.e. log($\rm SFR_{[OII]}/SFR_{MS}$) < $-$1.3, e.g. \citealt{Aird2019}). Hereafter we will refer to this subsample as low-SF Type II AGN.


Since our sample does not have H$\alpha$ and H$\beta$ within the observed spectral window, we could not compute the Balmer decrement and as a consequence did not correct the line luminosities for extinction. We explored the possible effect the extinction could have on the presence of the two populations.
 \cite{RosaGonzalez2002} show that the excess in the [OII]-based and UV-based SFR estimates is mainly due to an overestimation of the extinction resulting from the effect of underlying stellar Balmer absorptions in the measured emission line fluxes.
 Therefore they constructed unbiased SFR estimators, which statistically include the effect of underlying stellar Balmer absorptions in the measured emission line fluxes. 
 \cite{Kewley2004} found a strong correlation between the intrinsic [OII] luminosity and the color excess for the galaxies in Nearby Field Galaxies Survey, deriving a direct relation between intrinsic and observed [OII] luminosity, although they emphasize that the relation should not be blindly applied to other galaxies. We have tested what happens to the distribution shown in Fig \ref{fig:SFR_LOIII} applying either the extinction correction by \cite{Kewley2004}, eq. 18, or the recipe by \citealt{RosaGonzalez2002}. In both cases, we still find the observed separation between low- and high-SF AGN, and we can therefore conclude that the bimodality does not depend on the extinction.
 

In the following, we investigate whether the offset from the MS is related to the AGN power, by comparing it with the [OIII] luminosity. In Fig. \ref{fig:SFR_LOIII}, we show the median Log (SFR/SFR$\rm_{MS}$) in bins of [OIII] luminosity of the high and low-SF subsamples, with brown and orange circles respectively, where the errors bars indicate the 25th and 75th percentiles. We find a correlation between the relative offset of the SFR from the MS and the AGN power for both populations: at increasing AGN luminosities Type II AGN hosts tend to have higher SFR.
As a positive correlation exists between the [OII] and [OIII] luminosity, as a counter-check we performed the same analysis using the SFR as derived from the SED fitting, obtaining similar results but with a shallower slope, thus confirming the reliability of our findings (see stars in Fig. \ref{fig:SFR_LOIII}).

Previous works examined the connection between the SFR and AGN activity, with controversial results claiming strong to weak or absent relations (e.g. \citealt{Azadi2015}, \citealt{Chen2013}, \citealt{Stanley2015,Stanley2017}), \citealt{Harrison2012}. The origin of these discrepancies could be related to sample selection effects, methods of estimating the SFR and the AGN luminosity, as well as the number statistics of the sample (e.g. \citealt{Harrison2012}).
\citealt{Harrison2012} reported  that the SFR of z=1-3 AGN is independent of X-ray luminosity, used as indicator of AGN activity. This result is in contrast with that found by \cite{Page2012} and the authors suggest that a poor statistics is at least partially responsible for the disagreement at high luminosity between their work and that of \cite{Page2012}. \cite{Stanley2015} used 2000 X-ray detected AGN to investigate the SFR and AGN luminosity relation, in the redshift range 0.2$<$ z $<$2.5 and with X-ray luminosity 10$^{42}$<L$_{2-8kev}$ 10$^{45.5}$ erg s$^{-1}$. They used infrared SEDs decomposition (AGN+star formation components) to derive IR-based SFR and X-ray luminosity as a probe of AGN power, founding a broadly flat SFR-AGN luminosity relation at all redshifts and all the AGN luminosity investigated. They argue that the flat observed relation is probably due to short time-scale variations in AGN luminosity (probed by X-ray luminosity), which can wash out the long-term relationship between SFR and AGN activity. \cite{Masoura2021} found a positive correlation between the MS offset and the X-ray luminosity of a sample of X-ray selected type II AGN at 0.03<z<3.5.  \cite{Zhuang2020} analysed a sample of 5800 Type 1 and 7600 Type 2 AGN at z < 0.35 to study the star formation activity based on [OII] and [OIII] emission lines, finding a tight linear correlation between AGN luminosity (probed by [OIII] emission) and SFR. The [OIII] AGN indicator probes the AGN activity on longer timescales than X-ray luminosity, which traces the instantaneous AGN strength, and therefore the use of [OIII] may result in stronger correlation between SFR and AGN luminosity. 

The positive correlation found for SFR and AGN activity support the idea that the AGN and the star-formation activity in the host galaxy are sustained by a common fuelling mechanism, the large amounts of cold gas, and that the growths of the stellar mass and of the SMBH proceed concurrently (e.g. \citealt{Silverman2009}, \citealt{Santini2012}).


However at high stellar masses Type II AGN host galaxies show systematically lower SFR values. 
This could indicate that the process of AGN growth is linked to the process of star formation in AGN host galaxies (e.g. \citealt{Matsuoka2015}, \citealt{Mullaney2015}, \citealt{Shimizu2015}). The AGN could work against star formation, decreasing the gas reservoir through several ways such as mechanical heating and powerful outflows, moving the host galaxies of Type II AGN down to the quiescent locus in the SFR-M$\rm_{stellar}$ plane.





\subsection{Low-SF and high-SF Type II AGN properties}
\subsubsection{[OIII] line shape}


About 50\% of our type II AGN host galaxies are located below the MS. Their lower than expected SFRs might be evidence of on-going quenching. Powerful AGN radiation is often invoked as one of the main mechanisms to halt star formation by ejecting the gas necessary to fuel it. The ejected gas can be traced at all scales, and in various gas components: blue-shifted absorption lines from the accretion disc (i.e. ultra-fast outflows; e.g. \citealt{Tombesi2010}), blue-shifted emission line components produced by ionized gas in the broad line region, BLR (e.g., CIV; \citealt{Vietri2020} and references therein) and narrow line region, NLR (e.g., [OIII]; \citealt{Harrison2016}), and as blue-shifted lines produced by outflowing cold molecular gas on galactic scales (e.g., CO; \citealt{Polletta2011}, and many others).  Here, we investigate whether we find any evidence of AGN-driven outflowing gas by analysing the profile of the [OIII] emission line.

Each population (as defined in Sec. \ref{sec:populations}) is divided in [OIII] luminosity bins as done in Sec. \ref{sec:sfr_mass}.


We performed a median spectral stack, by using IRAF task {\it{scombine}}, resampling spectra to a rest-frame wavelength grid from  3520 \AA\ with a step size of 4.29 \AA, corresponding to the wavelength resolution at redshift 0.7, the mean redshift of the sources studied in this paper. 
We also normalized each spectrum to the continuum from 4500 \AA\ up to 4600 \AA, where the spectrum is free of strong emission and absorption lines. 

We analysed the line profile of the [OIII]$\lambda$4959,5007 doublet, H$\beta$ and [OII] doublet by fitting the lines with two models, considering a single and a double Gaussian to search for a possible second broad and shifted component, indicative of the presence of outflow. We used the same constraints as discussed in Sec. \ref{sec:analysis}.
We adopted the double Gaussian model as best-fit when it satisfies the Bayesian Information Criterion (BIC, \citealt{Schwarz1978}), which uses differences in $\chi^2$ that penalize models with more free parameters.
 For both models we estimated the BIC defined as $\chi^2$ + $k$ ln($N$), with $N$ the number of data points and $k$ the number of free parameters of the model. For each stacked spectrum we derived the $\Delta$BIC= BIC$\rm_1$ $-$ BIC$\rm_2$, where BIC$\rm_1$ and BIC$\rm_2$ are derived from the models with one and two Gaussian profiles, respectively. We favoured the fit with a single Gaussian profile when $\Delta$BIC < 10.



\begin{figure*}[]
\centering
  \includegraphics[width=0.45\textwidth]{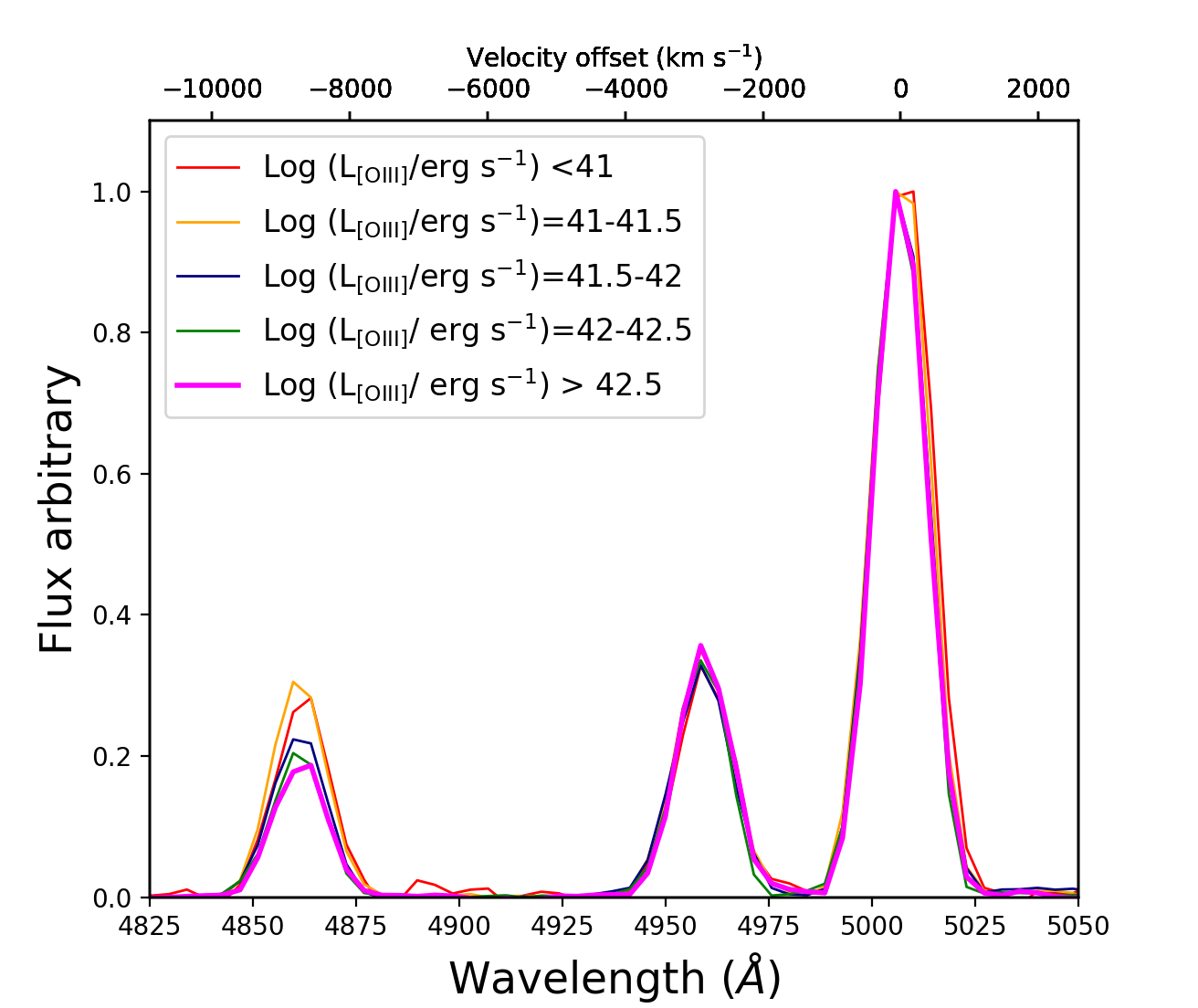}
  \includegraphics[width=0.47\textwidth]{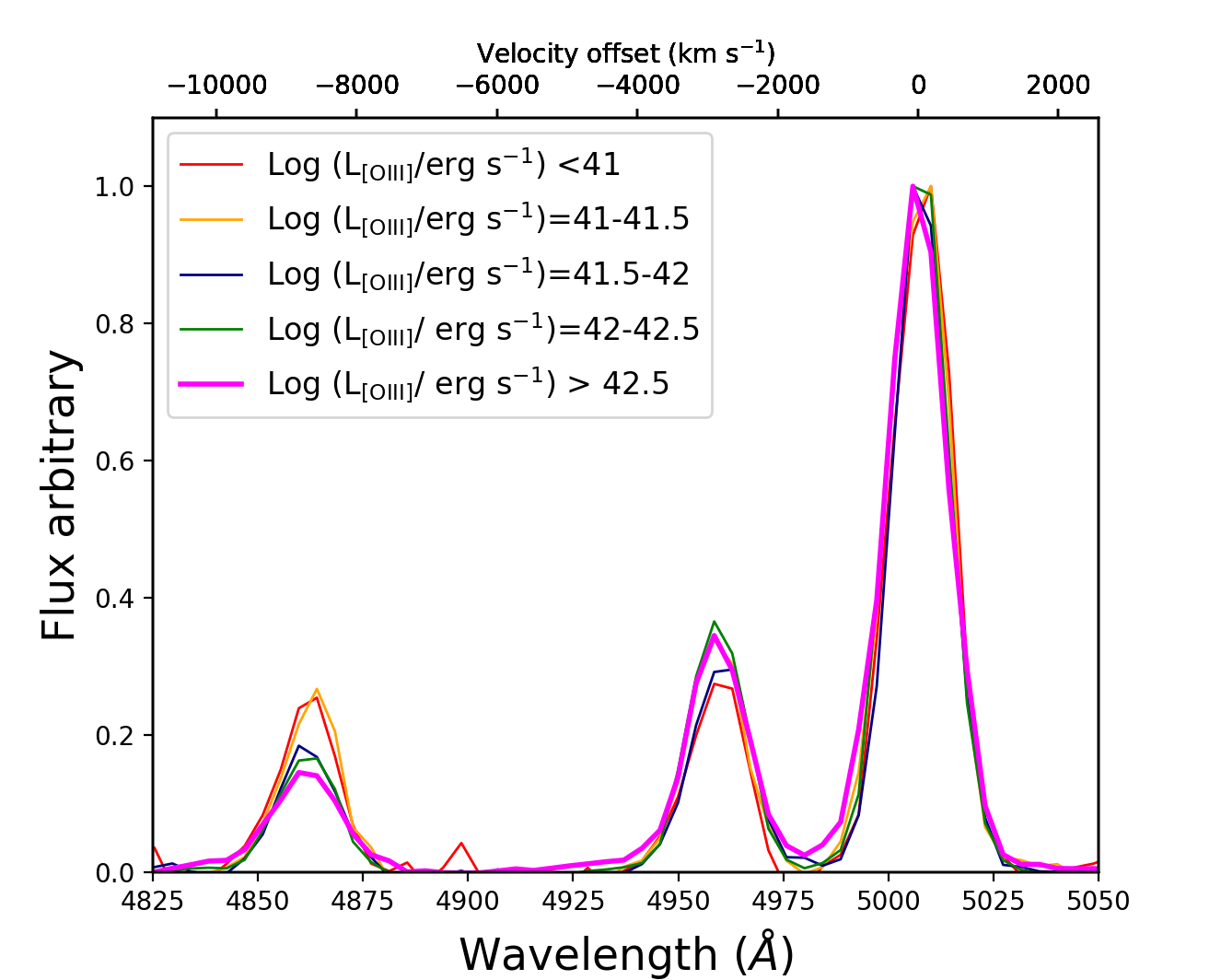}
  \includegraphics[width=0.45\textwidth]{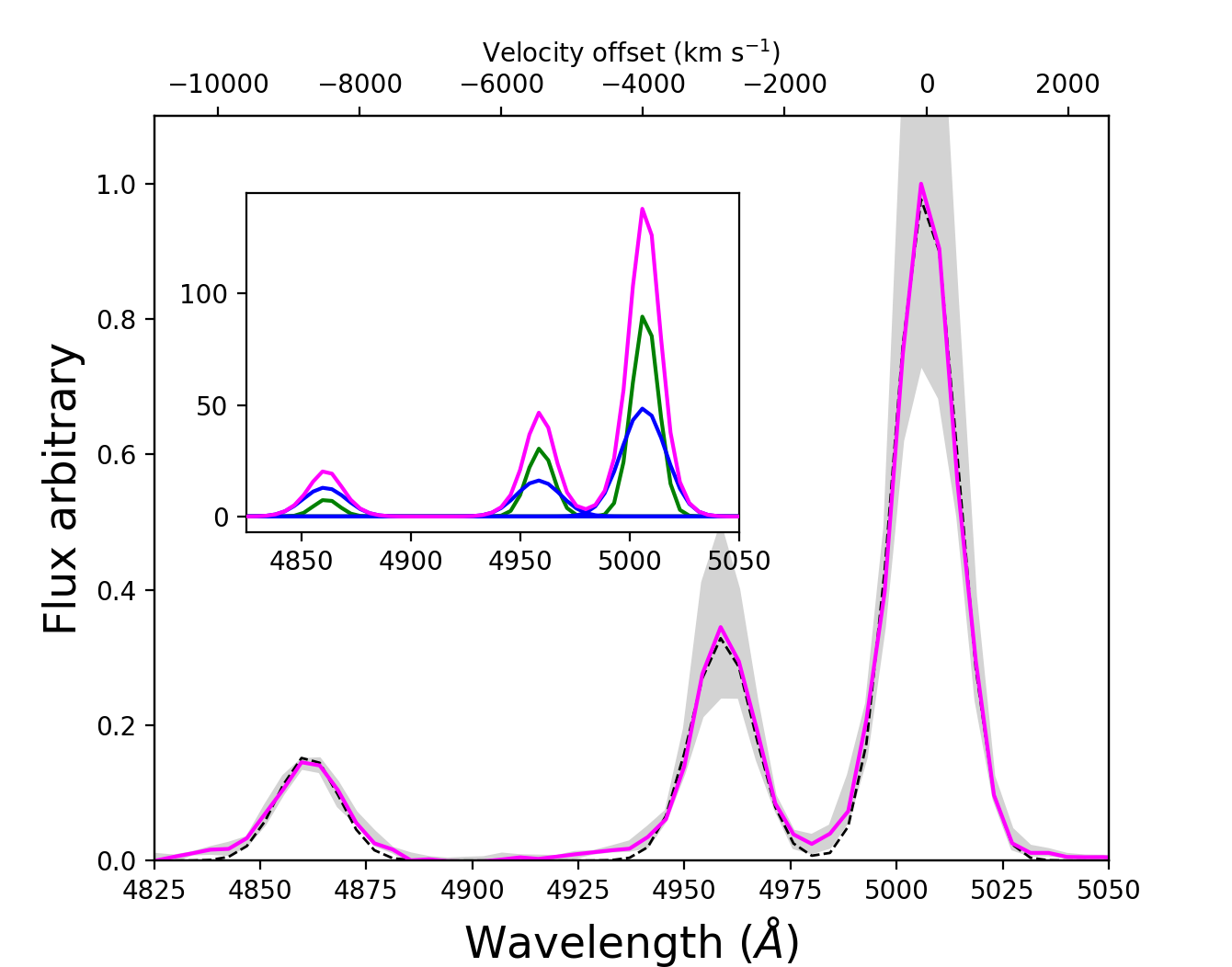}
 
\caption{Comparison of the stacked spectra, in the H$\beta$-[OIII] doublet lines wavelength range, in bins of [OIII]$\lambda$5007 luminosity above (left) and below (right) the line threshold as defined in Sec. \ref{sec:populations}. (lower panel) 
Stacked spectrum in the [OIII] luminosity bin Log (L$\rm_{[OIII]}/ erg\ s^{-1}$) > 42.5 of the low-SF Type II AGN with superimposed the 1$\sigma$ uncertainties estimated through a bootstrap resampling technique and the best fitting one-component Gaussian curve (black dashed line). The inset plot shows the best-fit double Gaussian model (magenta curve) and its line decomposition (green and blue Gaussian profiles refer to the narrow and broad best-fit components).}\label{fig:stack_spectra}

\end{figure*}


In Fig. \ref{fig:stack_spectra} we compared the spectral properties of the H$\beta$ and [OIII] doublet lines in each [OIII] luminosity bin and for each subsample. In all but the high luminosity bin of the low-SF Type II AGN hosts, the [OIII] line appears to be symmetrical as the one Gaussian preferred model and visual inspection suggest.

Only in the highest luminosity bin (Log (L$\rm_{[OIII]}/ erg\ s^{-1}$) > 42.5) of the low-SF population, there seems to be a hint of asymmetry in the [OIII] line profile. Fig. \ref{fig:stack_spectra} (bottom panel) shows the spectrum (magenta line) and the best fitting single component Gaussian (black dashed line). 
To rule out that such an excess is compatible with errors, we estimated the uncertainties on the stack through a bootstrap resampling technique, creating 1000 realizations of the AGN stack spectra with replacement, and derived the 1$\sigma$ uncertainties from 84th and 16th percentiles of the bootstrap distribution, shown as grey area in Fig. \ref{fig:stack_spectra} (bottom panel).   
Also in the fitting, a double Gaussian model is preferred, with a centroid of the second component nearly at systemic redshift and a FWHM of 1260 km/s, after correcting for instrumental broadening (see inset plot in Fig. \ref{fig:stack_spectra}, bottom panel.)

This finding is in agreement with previous results that reported an increasing outflow component at increasing luminosity (e.g. \citealt{Mullaney2013} ). The presence of outflowing gas in galaxies with stellar mass > 10$^{10}$ M$_{\odot}$ and their position in SFR-stellar mass plane is qualitatively consistent with the evolutionary scenario, where the AGN is capable of driving outflows that could regulate the star formation and the baryonic content of galaxies.

The line profile analysis indicates the presence of disturbed kinematics only in the high luminosity bin of the low-SF sample. Indeed we do not find a similar result for the high-SF sample at the highest luminosity bin. 
This could be either due to the absence of outflow in galaxies with stellar mass <10$^{10}$ M$\odot$ or
, considering the unified model, to the fact that the outflowing material should emerge in a direction perpendicular to the plane of the obscuring torus (i.e. to our line-of-sight), resulting in a small projected velocity of the outflow and/or in a symmetric line profile, which can explain the symmetric profiles found for most of the stacked VIMOS sample (see \citealt{Mullaney2013}, \citealt{Harrison2012}). In case the outflow signature is unresolved, it could be that the outflows are not quenching the star formation in these systems or that the timescale is actually longer than the stage at which these objects are seen.
 


Higher resolution spectroscopy becomes necessary to characterize subtle spectral features and disentangle between gravitational and non-gravitational motions, to get a deep insight on the AGN feedback in those systems.

\subsubsection{Black hole masses and Eddington ratios}
Previous studies have shown that there is a correlation between strong blue wings, large FWHM of line profiles originating in the NLR and Eddington ratio, which describes the accretion mechanism of an AGN (e.g. \citealt{Woo2016}).

The Eddington ratio is defined as:

\begin{equation}
\rm \lambda_{Edd}= \frac{L_{Bol}}{L_{Edd}} 
\end{equation}

where L$\rm_{Edd}$ (= 1.27 $\times\ 10^{38} M\rm_{BH}$, with M$\rm_{BH}$ indicating the BH mass) is the limit at which the outward radiation pressure from the accreting matter balances the inward gravitational pressure exerted by the BH, and $L\rm_{Bol}$ is the bolometric luminosity.

BH masses for Type I AGN are usually estimated indirectly by using the virial theorem, which links the BH mass to the BLR radius and the gas velocity dispersion. Considering the H$\alpha$ emission line, the single-epoch relation can be written as (\citealt{Baron2019}):
\begin{equation}
\rm \frac{M_{BH}}{M_{\odot}}=log \epsilon + 6.90 +0.54 \times log \frac{\lambda L_{\lambda} 5100}{10^{44} erg s^{-1}} +2.06 \times\ log \frac{FWHM_{H\alpha}^{BLR}}{10^3 km s^{-1}}
\end{equation}\label{eq:bh_mass}
with $\epsilon$ as the virial shape factor, $\lambda L_{\lambda} 5100$ the monochromatic AGN luminosity at 5100$\AA$ and FWHM$\rm_{H\alpha}^{BLR}$ the BLR component of the H$\alpha$ emission line.

However, Type II AGN are viewed edge-on, preventing us to see the BLR which is obscured by the presence of a dusty torus. Therefore in Type II AGN BH masses cannot be estimated by using the single-epoch mass determination which requires the view of BLR clouds. Indirect methods can be used as the well-known correlations between the BH mass and host galaxy bulge stellar mass or stellar velocity dispersion. However these relations are established 
for local inactive galaxies.
Recently \cite{Baron2019} find a correlation between the narrow L([OIII])/L(H$\beta$) line ratio and the width of the H$\alpha$ BLR component, linking the kinematics of the BLR clouds to the ionization state of the NLR as follows:

\begin{equation}
\rm log\frac{L_{[OIII]}^{narrow}}{L_{H\beta}^{narrow}}=0.58\pm0.07 \times log \frac{FWHM _{H_{\alpha}}^{BLR}}{km/s}-1.38\pm0.38\label{eq:ratio_fw}
\end{equation}

This power-law dependence holds for AGN-dominated systems with log([OIII]/H$\beta$)>0.55. We therefore derive the BLR H$\alpha$ FWHM component for the 79\% of our targets, exhibiting log([OIII]/H$\beta$)>0.55. 

For the continuum luminosity we rely on relation found by \cite{Baron2019}:

\begin{equation}
\rm Log \lambda L_{\lambda}5100 = 1.09 \times log L_{bol} -5.23
\end{equation}

and \cite{Heckman2004} for the bolometric luminosity, inferred from the [OIII] luminosity with no correction for dust extinction and applying a bolometric correction of 3500.  







\begin{figure}[]
\center
 \includegraphics[width=1\columnwidth]{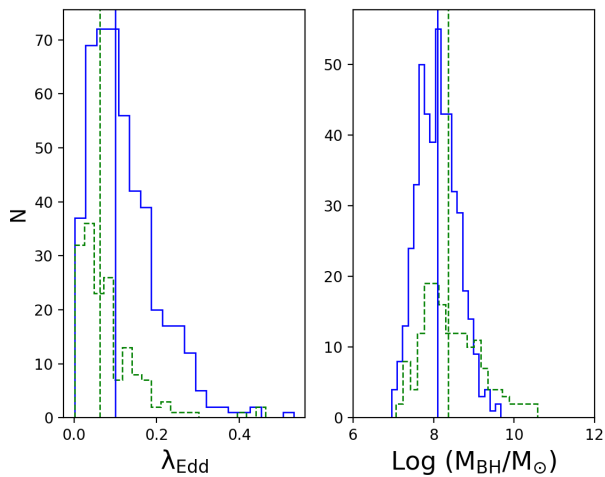}
 \caption{Eddington ratio (left) and black hole mass (right) distributions of Type II AGN divided into two subsamples, according to the classification of their host galaxies as high-SF (blue) and low-SF (green). 
 Solid and dashed lines represent the median value of the parameters.}\label{fig:BH_edd}
\end{figure}

We derive BH masses in the range $\sim$ 10$^{7-10}$ M$_{\odot}$ and \edd$\sim$10$^{-3}$-0.5, with a median value of $\sim$0.08, consistent with what is found in other AGN samples (e.g. \citealt{Lamastra2009})
We now explore if there is an indication for a variation of BH mass and Eddington ratio distribution between the Type II AGN groups in Fig. \ref{fig:BH_edd} (parameter values for high-SF galaxies are shown as blue solid lines and for low-SF sample as green dashed vertical lines).
Despite the wide range spanned, differences between BH mass and \edd\ distributions are discernible. The median values for each parameter are shown as vertical lines in corresponding colors and line styles. The median value of $\lambda_{Edd}$ derived for the high-SF galaxies is slightly higher than that derived for low-SF galaxies. Furthermore these galaxies are less massive and have in general lower BH mass than low-SF galaxies. On the contrary, this latter galaxy sample shows larger values of BH mass, the distribution of which points towards higher BH mass values, as shown by the M$\rm_{BH}$ distribution in Fig. \ref{fig:BH_edd}. 

To assess difference between the distributions, we compute the Kolmogorov-Smirnov (K-S) test for the $\lambda_{Edd}$ and BH mass distributions of the two groups of Type II AGN. The null-hypothesis is that the two samples are drawn from the same parent population. The K-S test was performed by using the python routine  {\tt{scipy.stat.ks\_2samp}}. For the $\lambda_{Edd}$ distributions we compute a statistic of 0.3 and p-value = 2.6 10$^{-8}$ and for BH mass distributions a statistic of 0.2 and p-value = 4.5 10$^{-6}$. The very low probability value excludes that the two groups of Type II AGN are drawn from the same parent population,
 suggesting that the difference between these groups also relates to the AGN activity.

It is instructive to compare these findings with what is found in the local Universe. \cite{Kauffmann2009} examined the dependence of the distribution of the Eddington ratio on the star formation history of SDSS Type II AGN. They found that the Eddington ratio distributions can be different for actively star-forming and passive host galaxies. Specifically they divided these two populations according to the break index D(4000) (\citealt{Balogh1999}), a useful diagnostic of the recent star formation history in these systems.  The star-forming host galaxies show a log-normal distribution of Eddington ratios peaking at a few percent of Eddington, independent of black hole masses. This regime dominates the growth of BH with mass <10$^8$ M$_{\odot}$. While the passive galaxies (D4000 >1.7) show a power-law distribution of Eddington ratios with its amplitude decreasing with increasing black hole mass. This regime dominates the growth of BH with mass >10$^8$ M$_{\odot}$.
This finding is in line with our evidence of different accretion rates distribution for high and low-SF Type II AGN host galaxies.

Finally, although the estimated Eddington ratios have large uncertainties, we did not find a clear dependence of the broadening of the [OIII] line with the Eddington ratio, since the only evidence of outflowing gas is found in the [OIII]-luminous low-SF galaxies, which are accreting at $\sim$5\% of the Eddington rate, a lower rate than that of high-SF Type II AGN subsample, which show on average Eddington ratios of $\sim$0.10. This could be interpreted as a final decay phase of the AGN activity in the [OIII]-luminous low-SF galaxies, where the outflowing gas persists but the AGN feeding mechanism is fading as well as the star formation activity, likely due to AGN feedback.

\section{Summary and conclusions}\label{sec:summary}
In this paper we have used the VVDS and VIPERS optical spectroscopic surveys, carried out using the VIMOS spectrograph, to select and investigate the properties of those galaxies hosting an AGN at redshift 0.5<z<0.9. We have analysed the emission line properties of the [OIII] doublet, H$\beta$ and [OII] doublet and adopted the blue diagram of \cite{Lamareille2010} to distinguish Type II AGN from star-forming galaxies and LINERS, through emission line ratios [OIII]/H$\beta$ vs [OII]/H$\beta$.

Our main findings can be summarized as follows:

(i) The masses of the host galaxies range from 10$^{8}$ to 10$^{12}$ M$_{\odot}$, with a median value of 10$^{9.5}$ M$_{\odot}$ and span a star-formation rates range of 0.01-38 M$_{\odot}$/yr. The VIMOS sample with stellar mass < 10$^{10}$ M$_{\odot}$ mostly resides on the  star-forming MS locus, as defined by \cite{Schreiber2015}, with a fraction of sources ($\sim$20\%) between the MS and quiescent region having stellar mass >  10$^{10}$ M$_{\odot}$, indicating reduced level of star formation.

(ii) We find a bimodality in the SFR MS offset-AGN power plane (probed by the [OIII] luminosity), ascribing to two different populations in the VIMOS sample.
We divide our type II AGN sample in two groups according to the properties of their host galaxies, the high-SF one with stellar mass < 10$^{10}$ M$_{\odot}$, occupying the star-forming MS region and the low-SF one with levels of star formation between the MS and quiescent locus, and even lower, and stellar mass > 10$^{10}$ M$_{\odot}$. For both populations a positive correlation exists between the distance from the MS and the AGN power, which could reflect the available amount of gas which both triggers star formation and fuels the AGN activity. Despite this positive correlation, lower level of star formation rates are found in low-SF Type II galaxies.

(iii) AGN feedback may be responsible for reducing the supply of cold gas in host-galaxies at least for AGN luminous systems. Indeed,  for the [OIII]-luminous low-SF galaxies we found a hint of outflowing gas, as probed by the asymmetric [OIII] line profile, which could be connected with the low SFR content found, possibly due to the effect of AGN acting on the ISM, expelling a certain amount of gas. These massive low-SF galaxies seem to be at their final AGN stage as indicated by their average Eddington ratio value ($\sim$5\% of Eddington rate).


\begin{acknowledgements}
We thank the anonymous referee for the helpful comments. GV acknowledges financial support from Premiale 2015 MITic (PI: B. Garilli). AP and KM acknowledge support from the Polish National Science Centre under grants: UMO-2018/30/M/ST9/00757 and  UMO-2018/30/E/ST9/00082. GM acknowledges support from ST/P006744/1. The results published have been funded by the European Union's Horizon 2020 research and innovation programme under the Maria Skłodowska-Curie (grant agreement No 754510), the National Science Centre of Poland (grant UMO-2016/23/N/ST9/02963) and by the Spanish Ministry of Science and Innovation through Juan de la Cierva-formacion program (reference FJC2018-038792-I). Based on observations made with ESO Telescopes at the La Silla or Paranal Observatories under programme ID(s) 182.A-0886(H), 182.A-0886(N), 182.A-0886(K), 182.A-0886(R), 182.A-0886(G), 182.A-0886(C), 182.A-0886(B), 182.A-0886(O), 182.A-0886(P), 
182.A-0886(J), 182.A-0886(D), 182.A-0886(I), 182.A-0886(Q), 60.A-9157(B). Based on data obtained with the European Southern Observatory Very Large Telescope, Paranal, Chile, under Large Programmes 070.A-9007 and 177.A-0837.
\end{acknowledgements}

\bibliographystyle{aa} 
\bibliography{bib} 

\begin{thebibliography}{123}
\expandafter\ifx\csname natexlab\endcsname\relax\def\natexlab#1{#1}\fi

\bibitem[{{Aird} {et~al.}(2019){Aird}, {Coil}, \& {Georgakakis}}]{Aird2019}
{Aird}, J., {Coil}, A.~L., \& {Georgakakis}, A. 2019, \mnras, 484, 4360

\bibitem[{{Aird} {et~al.}(2015){Aird}, {Coil}, {Georgakakis}, {Nandra},
  {Barro}, \& {P{\'e}rez-Gonz{\'a}lez}}]{Aird2015}
{Aird}, J., {Coil}, A.~L., {Georgakakis}, A., {et~al.} 2015, \mnras, 451, 1892

\bibitem[{{Alexander} \& {Hickox}(2012)}]{Alexander2012}
{Alexander}, D.~M. \& {Hickox}, R.~C. 2012, \nar, 56, 93

\bibitem[{{Antonucci}(1993)}]{Antonucci1993}
{Antonucci}, R. 1993, \araa, 31, 473

\bibitem[{{Arnouts} {et~al.}(2005){Arnouts}, {Schiminovich}, {Ilbert},
  {Tresse}, {Milliard}, {Treyer}, {Bardelli}, {Budavari}, {Wyder}, {Zucca}, {Le
  F{\`e}vre}, {Martin}, {Vettolani}, {Adami}, {Arnaboldi}, {Barlow}, {Bianchi},
  {Bolzonella}, {Bottini}, {Byun}, {Cappi}, {Charlot}, {Contini}, {Donas},
  {Forster}, {Foucaud}, {Franzetti}, {Friedman}, {Garilli}, {Gavignaud},
  {Guzzo}, {Heckman}, {Hoopes}, {Iovino}, {Jelinsky}, {Le Brun}, {Lee},
  {Maccagni}, {Madore}, {Malina}, {Marano}, {Marinoni}, {McCracken}, {Mazure},
  {Meneux}, {Merighi}, {Morrissey}, {Neff}, {Paltani}, {Pell{\`o}}, {Picat},
  {Pollo}, {Pozzetti}, {Radovich}, {Rich}, {Scaramella}, {Scodeggio},
  {Seibert}, {Siegmund}, {Small}, {Szalay}, {Welsh}, {Xu}, {Zamorani}, \&
  {Zanichelli}}]{Arnouts2005}
{Arnouts}, S., {Schiminovich}, D., {Ilbert}, O., {et~al.} 2005, \apjl, 619, L43

\bibitem[{{Azadi} {et~al.}(2015){Azadi}, {Aird}, {Coil}, {Moustakas}, {Mendez},
  {Blanton}, {Cool}, {Eisenstein}, {Wong}, \& {Zhu}}]{Azadi2015}
{Azadi}, M., {Aird}, J., {Coil}, A.~L., {et~al.} 2015, \apj, 806, 187

\bibitem[{{Baldry} {et~al.}(2004){Baldry}, {Glazebrook}, {Brinkmann},
  {Ivezi{\'c}}, {Lupton}, {Nichol}, \& {Szalay}}]{Baldry2004}
{Baldry}, I.~K., {Glazebrook}, K., {Brinkmann}, J., {et~al.} 2004, \apj, 600,
  681

\bibitem[{{Baldwin} {et~al.}(1981){Baldwin}, {Phillips}, \&
  {Terlevich}}]{Baldwin1981}
{Baldwin}, J.~A., {Phillips}, M.~M., \& {Terlevich}, R. 1981, \pasp, 93, 5

\bibitem[{{Balogh} {et~al.}(1999){Balogh}, {Morris}, {Yee}, {Carlberg}, \&
  {Ellingson}}]{Balogh1999}
{Balogh}, M.~L., {Morris}, S.~L., {Yee}, H.~K.~C., {Carlberg}, R.~G., \&
  {Ellingson}, E. 1999, \apj, 527, 54

\bibitem[{{Baron} \& {M{\'e}nard}(2019)}]{Baron2019}
{Baron}, D. \& {M{\'e}nard}, B. 2019, \mnras, 487, 3404

\bibitem[{{Bielby} {et~al.}(2012){Bielby}, {Hudelot}, {McCracken}, {Ilbert},
  {Daddi}, {Le F{\`e}vre}, {Gonzalez-Perez}, {Kneib}, {Marmo}, {Mellier},
  {Salvato}, {Sanders}, \& {Willott}}]{Bielby2012}
{Bielby}, R., {Hudelot}, P., {McCracken}, H.~J., {et~al.} 2012, \aap, 545, A23

\bibitem[{{Blanton} {et~al.}(2003){Blanton}, {Hogg}, {Bahcall}, {Baldry},
  {Brinkmann}, {Csabai}, {Eisenstein}, {Fukugita}, {Gunn}, {Ivezi{\'c}},
  {Lamb}, {Lupton}, {Loveday}, {Munn}, {Nichol}, {Okamura}, {Schlegel},
  {Shimasaku}, {Strauss}, {Vogeley}, \& {Weinberg}}]{Blanton2003}
{Blanton}, M.~R., {Hogg}, D.~W., {Bahcall}, N.~A., {et~al.} 2003, \apj, 594,
  186

\bibitem[{{Bluck} {et~al.}(2014){Bluck}, {Mendel}, {Ellison}, {Moreno},
  {Simard}, {Patton}, \& {Starkenburg}}]{Bluck2014}
{Bluck}, A. F.~L., {Mendel}, J.~T., {Ellison}, S.~L., {et~al.} 2014, \mnras,
  441, 599

\bibitem[{{Bongiorno} {et~al.}(2012){Bongiorno}, {Merloni}, {Brusa},
  {Magnelli}, {Salvato}, {Mignoli}, {Zamorani}, {Fiore}, {Rosario}, {Mainieri},
  {Hao}, {Comastri}, {Vignali}, {Balestra}, {Bardelli}, {Berta}, {Civano},
  {Kampczyk}, {Le Floc'h}, {Lusso}, {Lutz}, {Pozzetti}, {Pozzi}, {Riguccini},
  {Shankar}, \& {Silverman}}]{Bongiorno2012}
{Bongiorno}, A., {Merloni}, A., {Brusa}, M., {et~al.} 2012, \mnras, 427, 3103

\bibitem[{{Boquien} {et~al.}(2019){Boquien}, {Burgarella}, {Roehlly}, {Buat},
  {Ciesla}, {Corre}, {Inoue}, \& {Salas}}]{Boquien2019}
{Boquien}, M., {Burgarella}, D., {Roehlly}, Y., {et~al.} 2019, \aap, 622, A103

\bibitem[{{Bruzual} \& {Charlot}(2003)}]{Bruzual2003}
{Bruzual}, G. \& {Charlot}, S. 2003, \mnras, 344, 1000

\bibitem[{{Calzetti} {et~al.}(2000){Calzetti}, {Armus}, {Bohlin}, {Kinney},
  {Koornneef}, \& {Storchi-Bergmann}}]{Calzetti2000}
{Calzetti}, D., {Armus}, L., {Bohlin}, R.~C., {et~al.} 2000, \apj, 533, 682

\bibitem[{{Cappellari}(2012)}]{Cappellari2012}
{Cappellari}, M. 2012, {pPXF: Penalized Pixel-Fitting stellar kinematics
  extraction}

\bibitem[{{Cattaneo} {et~al.}(2009){Cattaneo}, {Faber}, {Binney}, {Dekel},
  {Kormendy}, {Mushotzky}, {Babul}, {Best}, {Br{\"u}ggen}, {Fabian}, {Frenk},
  {Khalatyan}, {Netzer}, {Mahdavi}, {Silk}, {Steinmetz}, \&
  {Wisotzki}}]{Cattaneo2009}
{Cattaneo}, A., {Faber}, S.~M., {Binney}, J., {et~al.} 2009, \nat, 460, 213

\bibitem[{{Chen} {et~al.}(2013){Chen}, {Hickox}, {Alberts}, {Brodwin}, {Jones},
  {Murray}, {Alexander}, {Assef}, {Brown}, {Dey}, {Forman}, {Gorjian},
  {Goulding}, {Le Floc'h}, {Jannuzi}, {Mullaney}, \& {Pope}}]{Chen2013}
{Chen}, C.-T.~J., {Hickox}, R.~C., {Alberts}, S., {et~al.} 2013, \apj, 773, 3

\bibitem[{{Ciesla} {et~al.}(2015){Ciesla}, {Charmandaris}, {Georgakakis},
  {Bernhard}, {Mitchell}, {Buat}, {Elbaz}, {LeFloc'h}, {Lacey}, {Magdis}, \&
  {Xilouris}}]{Ciesla2015}
{Ciesla}, L., {Charmandaris}, V., {Georgakakis}, A., {et~al.} 2015, \aap, 576,
  A10

\bibitem[{{Cowie} {et~al.}(1996){Cowie}, {Songaila}, {Hu}, \&
  {Cohen}}]{Cowie1996}
{Cowie}, L.~L., {Songaila}, A., {Hu}, E.~M., \& {Cohen}, J.~G. 1996, \aj, 112,
  839

\bibitem[{{Croton} {et~al.}(2006){Croton}, {Springel}, {White}, {De Lucia},
  {Frenk}, {Gao}, {Jenkins}, {Kauffmann}, {Navarro}, \& {Yoshida}}]{Croton2006}
{Croton}, D.~J., {Springel}, V., {White}, S. D.~M., {et~al.} 2006, \mnras, 365,
  11

\bibitem[{{Cuillandre} {et~al.}(2012){Cuillandre}, {Withington}, {Hudelot},
  {Goranova}, {McCracken}, {Magnard}, {Mellier}, {Regnault}, {B{\'e}toule},
  {Aussel}, {Kavelaars}, {Fernique}, {Bonnarel}, {Ochsenbein}, \&
  {Ilbert}}]{Cuillandre2012}
{Cuillandre}, J.-C.~J., {Withington}, K., {Hudelot}, P., {et~al.} 2012, in
  Society of Photo-Optical Instrumentation Engineers (SPIE) Conference Series,
  Vol. 8448, Observatory Operations: Strategies, Processes, and Systems IV, ed.
  A.~B. {Peck}, R.~L. {Seaman}, \& F.~{Comeron}, 84480M

\bibitem[{{Daddi} {et~al.}(2007){Daddi}, {Dickinson}, {Morrison}, {Chary},
  {Cimatti}, {Elbaz}, {Frayer}, {Renzini}, {Pope}, {Alexander}, {Bauer},
  {Giavalisco}, {Huynh}, {Kurk}, \& {Mignoli}}]{Daddi2007}
{Daddi}, E., {Dickinson}, M., {Morrison}, G., {et~al.} 2007, \apj, 670, 156

\bibitem[{{Dale} {et~al.}(2014){Dale}, {Helou}, {Magdis}, {Armus},
  {D{\'\i}az-Santos}, \& {Shi}}]{Dale2014}
{Dale}, D.~A., {Helou}, G., {Magdis}, G.~E., {et~al.} 2014, \apj, 784, 83

\bibitem[{{Di Matteo} {et~al.}(2005){Di Matteo}, {Springel}, \&
  {Hernquist}}]{DiMatteo2005}
{Di Matteo}, T., {Springel}, V., \& {Hernquist}, L. 2005, \nat, 433, 604

\bibitem[{{Elbaz} {et~al.}(2007){Elbaz}, {Daddi}, {Le Borgne}, {Dickinson},
  {Alexander}, {Chary}, {Starck}, {Brandt}, {Kitzbichler}, {MacDonald},
  {Nonino}, {Popesso}, {Stern}, \& {Vanzella}}]{Elbaz2007}
{Elbaz}, D., {Daddi}, E., {Le Borgne}, D., {et~al.} 2007, \aap, 468, 33

\bibitem[{{Faucher-Gigu{\`e}re} \& {Quataert}(2012)}]{Faucher2012}
{Faucher-Gigu{\`e}re}, C.-A. \& {Quataert}, E. 2012, \mnras, 425, 605

\bibitem[{{Ferrarese} \& {Merritt}(2000)}]{Ferrarese2000}
{Ferrarese}, L. \& {Merritt}, D. 2000, \apjl, 539, L9

\bibitem[{{Fritz} {et~al.}(2014){Fritz}, {Scodeggio}, {Ilbert}, {Bolzonella},
  {Davidzon}, {Coupon}, {Garilli}, {Guzzo}, {Zamorani}, {Abbas}, {Adami},
  {Arnouts}, {Bel}, {Bottini}, {Branchini}, {Cappi}, {Cucciati}, {De Lucia},
  {de la Torre}, {Franzetti}, {Fumana}, {Granett}, {Iovino}, {Krywult}, {Le
  Brun}, {Le F{\`e}vre}, {Maccagni}, {Ma{\l}ek}, {Marulli}, {McCracken},
  {Paioro}, {Polletta}, {Pollo}, {Schlagenhaufer}, {Tasca}, {Tojeiro},
  {Vergani}, {Zanichelli}, {Burden}, {Di Porto}, {Marchetti}, {Marinoni},
  {Mellier}, {Moscardini}, {Nichol}, {Peacock}, {Percival}, {Phleps}, \&
  {Wolk}}]{Fritz2014}
{Fritz}, A., {Scodeggio}, M., {Ilbert}, O., {et~al.} 2014, \aap, 563, A92

\bibitem[{{Fritz} {et~al.}(2006){Fritz}, {Franceschini}, \&
  {Hatziminaoglou}}]{Fritz2006}
{Fritz}, J., {Franceschini}, A., \& {Hatziminaoglou}, E. 2006, \mnras, 366, 767

\bibitem[{{Gabor} {et~al.}(2010){Gabor}, {Dav{\'e}}, {Finlator}, \&
  {Oppenheimer}}]{Gabor2010}
{Gabor}, J.~M., {Dav{\'e}}, R., {Finlator}, K., \& {Oppenheimer}, B.~D. 2010,
  \mnras, 407, 749

\bibitem[{{Gadotti}(2009)}]{Gadotti2009}
{Gadotti}, D.~A. 2009, \mnras, 393, 1531

\bibitem[{{Garilli} {et~al.}(2014){Garilli}, {Guzzo}, {Scodeggio},
  {Bolzonella}, {Abbas}, {Adami}, {Arnouts}, {Bel}, {Bottini}, {Branchini},
  {Cappi}, {Coupon}, {Cucciati}, {Davidzon}, {De Lucia}, {de la Torre},
  {Franzetti}, {Fritz}, {Fumana}, {Granett}, {Ilbert}, {Iovino}, {Krywult}, {Le
  Brun}, {Le F{\`e}vre}, {Maccagni}, {Ma{\l}ek}, {Marulli}, {McCracken},
  {Paioro}, {Polletta}, {Pollo}, {Schlagenhaufer}, {Tasca}, {Tojeiro},
  {Vergani}, {Zamorani}, {Zanichelli}, {Burden}, {Di Porto}, {Marchetti},
  {Marinoni}, {Mellier}, {Moscardini}, {Nichol}, {Peacock}, {Percival},
  {Phleps}, \& {Wolk}}]{Garilli2014}
{Garilli}, B., {Guzzo}, L., {Scodeggio}, M., {et~al.} 2014, \aap, 562, A23

\bibitem[{{Gebhardt} {et~al.}(2000){Gebhardt}, {Bender}, {Bower}, {Dressler},
  {Faber}, {Filippenko}, {Green}, {Grillmair}, {Ho}, {Kormendy}, {Lauer},
  {Magorrian}, {Pinkney}, {Richstone}, \& {Tremaine}}]{Gebhardt2000}
{Gebhardt}, K., {Bender}, R., {Bower}, G., {et~al.} 2000, \apjl, 539, L13

\bibitem[{{Giovannoli} {et~al.}(2011){Giovannoli}, {Buat}, {Noll},
  {Burgarella}, \& {Magnelli}}]{Giovannoli2011}
{Giovannoli}, E., {Buat}, V., {Noll}, S., {Burgarella}, D., \& {Magnelli}, B.
  2011, \aap, 525, A150

\bibitem[{{G{\"u}ltekin} {et~al.}(2009){G{\"u}ltekin}, {Richstone}, {Gebhardt},
  {Lauer}, {Tremaine}, {Aller}, {Bender}, {Dressler}, {Faber}, {Filippenko},
  {Green}, {Ho}, {Kormendy}, {Magorrian}, {Pinkney}, \&
  {Siopis}}]{Gultekin2009}
{G{\"u}ltekin}, K., {Richstone}, D.~O., {Gebhardt}, K., {et~al.} 2009, \apj,
  698, 198

\bibitem[{{Guzzo} {et~al.}(2014){Guzzo}, {Scodeggio}, {Garilli}, {Granett},
  {Fritz}, {Abbas}, {Adami}, {Arnouts}, {Bel}, {Bolzonella}, {Bottini},
  {Branchini}, {Cappi}, {Coupon}, {Cucciati}, {Davidzon}, {De Lucia}, {de la
  Torre}, {Franzetti}, {Fumana}, {Hudelot}, {Ilbert}, {Iovino}, {Krywult}, {Le
  Brun}, {Le F{\`e}vre}, {Maccagni}, {Ma{\l}ek}, {Marulli}, {McCracken},
  {Paioro}, {Peacock}, {Polletta}, {Pollo}, {Schlagenhaufer}, {Tasca},
  {Tojeiro}, {Vergani}, {Zamorani}, {Zanichelli}, {Burden}, {Di Porto},
  {Marchetti}, {Marinoni}, {Mellier}, {Moscardini}, {Nichol}, {Percival},
  {Phleps}, \& {Wolk}}]{Guzzo2014}
{Guzzo}, L., {Scodeggio}, M., {Garilli}, B., {et~al.} 2014, \aap, 566, A108

\bibitem[{{Harrison}(2017)}]{Harrison2017}
{Harrison}, C.~M. 2017, Nature Astronomy, 1, 0165

\bibitem[{{Harrison} {et~al.}(2016){Harrison}, {Alexander}, {Mullaney},
  {Stott}, {Swinbank}, {Arumugam}, {Bauer}, {Bower}, {Bunker}, \&
  {Sharples}}]{Harrison2016}
{Harrison}, C.~M., {Alexander}, D.~M., {Mullaney}, J.~R., {et~al.} 2016,
  \mnras, 456, 1195

\bibitem[{{Harrison} {et~al.}(2012){Harrison}, {Alexander}, {Swinbank},
  {Smail}, {Alaghband-Zadeh}, {Bauer}, {Chapman}, {Del Moro}, {Hickox},
  {Ivison}, {Men{\'e}ndez-Delmestre}, {Mullaney}, \& {Nesvadba}}]{Harrison2012}
{Harrison}, C.~M., {Alexander}, D.~M., {Swinbank}, A.~M., {et~al.} 2012,
  \mnras, 426, 1073

\bibitem[{{Hasinger} {et~al.}(2005){Hasinger}, {Miyaji}, \&
  {Schmidt}}]{Hasinger2005}
{Hasinger}, G., {Miyaji}, T., \& {Schmidt}, M. 2005, \aap, 441, 417

\bibitem[{{Heckman} {et~al.}(2004){Heckman}, {Kauffmann}, {Brinchmann},
  {Charlot}, {Tremonti}, \& {White}}]{Heckman2004}
{Heckman}, T.~M., {Kauffmann}, G., {Brinchmann}, J., {et~al.} 2004, \apj, 613,
  109

\bibitem[{{Hickox} {et~al.}(2014){Hickox}, {Mullaney}, {Alexander}, {Chen},
  {Civano}, {Goulding}, \& {Hainline}}]{Hickox2014}
{Hickox}, R.~C., {Mullaney}, J.~R., {Alexander}, D.~M., {et~al.} 2014, \apj,
  782, 9

\bibitem[{{Ho}(2005)}]{Ho2005}
{Ho}, L.~C. 2005, \apj, 629, 680

\bibitem[{{Hopkins} {et~al.}(2003){Hopkins}, {Miller}, {Nichol}, {Connolly},
  {Bernardi}, {G{\'o}mez}, {Goto}, {Tremonti}, {Brinkmann}, {Ivezi{\'c}}, \&
  {Lamb}}]{Hopkins2003}
{Hopkins}, A.~M., {Miller}, C.~J., {Nichol}, R.~C., {et~al.} 2003, \apj, 599,
  971

\bibitem[{{Hopkins} {et~al.}(2007){Hopkins}, {Richards}, \&
  {Hernquist}}]{Hopkins2007}
{Hopkins}, P.~F., {Richards}, G.~T., \& {Hernquist}, L. 2007, \apj, 654, 731

\bibitem[{{Hopkins} {et~al.}(2016){Hopkins}, {Torrey}, {Faucher-Gigu{\`e}re},
  {Quataert}, \& {Murray}}]{Hopkins2016}
{Hopkins}, P.~F., {Torrey}, P., {Faucher-Gigu{\`e}re}, C.-A., {Quataert}, E.,
  \& {Murray}, N. 2016, \mnras, 458, 816

\bibitem[{{Kauffmann} \& {Haehnelt}(2000)}]{Kauffmann2000}
{Kauffmann}, G. \& {Haehnelt}, M. 2000, \mnras, 311, 576

\bibitem[{{Kauffmann} \& {Heckman}(2009)}]{Kauffmann2009}
{Kauffmann}, G. \& {Heckman}, T.~M. 2009, \mnras, 397, 135

\bibitem[{{Kauffmann} {et~al.}(2003{\natexlab{a}}){Kauffmann}, {Heckman},
  {Tremonti}, {Brinchmann}, {Charlot}, {White}, {Ridgway}, {Brinkmann},
  {Fukugita}, {Hall}, {Ivezi{\'c}}, {Richards}, \&
  {Schneider}}]{Kauffmann2003a}
{Kauffmann}, G., {Heckman}, T.~M., {Tremonti}, C., {et~al.} 2003{\natexlab{a}},
  \mnras, 346, 1055

\bibitem[{{Kauffmann} {et~al.}(2003{\natexlab{b}}){Kauffmann}, {Heckman},
  {White}, {Charlot}, {Tremonti}, {Brinchmann}, {Bruzual}, {Peng}, {Seibert},
  {Bernardi}, {Blanton}, {Brinkmann}, {Castander}, {Cs{\'a}bai}, {Fukugita},
  {Ivezic}, {Munn}, {Nichol}, {Padmanabhan}, {Thakar}, {Weinberg}, \&
  {York}}]{Kauffmann2003c}
{Kauffmann}, G., {Heckman}, T.~M., {White}, S. D.~M., {et~al.}
  2003{\natexlab{b}}, \mnras, 341, 33

\bibitem[{{Kauffmann} {et~al.}(2003{\natexlab{c}}){Kauffmann}, {Heckman},
  {White}, {Charlot}, {Tremonti}, {Peng}, {Seibert}, {Brinkmann}, {Nichol},
  {SubbaRao}, \& {York}}]{Kauffmann2003b}
{Kauffmann}, G., {Heckman}, T.~M., {White}, S. D.~M., {et~al.}
  2003{\natexlab{c}}, \mnras, 341, 54

\bibitem[{{Kennicutt}(1998)}]{Kennicutt1998}
{Kennicutt}, Robert~C., J. 1998, \araa, 36, 189

\bibitem[{{Kewley} {et~al.}(2004){Kewley}, {Geller}, \& {Jansen}}]{Kewley2004}
{Kewley}, L.~J., {Geller}, M.~J., \& {Jansen}, R.~A. 2004, \aj, 127, 2002

\bibitem[{{Kim} {et~al.}(2006){Kim}, {Ho}, \& {Im}}]{Kim2006}
{Kim}, M., {Ho}, L.~C., \& {Im}, M. 2006, \apj, 642, 702

\bibitem[{Kormendy \& Ho(2013)}]{Kormendy2013}
Kormendy, J. \& Ho, L.~C. 2013, Annual Review of Astronomy and Astrophysics,
  51, 511

\bibitem[{{Lamareille}(2010)}]{Lamareille2010}
{Lamareille}, F. 2010, \aap, 509, A53

\bibitem[{{Lamareille} {et~al.}(2009){Lamareille}, {Brinchmann}, {Contini},
  {Walcher}, {Charlot}, {P{\'e}rez-Montero}, {Zamorani}, {Pozzetti},
  {Bolzonella}, {Garilli}, {Paltani}, {Bongiorno}, {Le F{\`e}vre}, {Bottini},
  {Le Brun}, {Maccagni}, {Scaramella}, {Scodeggio}, {Tresse}, {Vettolani},
  {Zanichelli}, {Adami}, {Arnouts}, {Bardelli}, {Cappi}, {Ciliegi}, {Foucaud},
  {Franzetti}, {Gavignaud}, {Guzzo}, {Ilbert}, {Iovino}, {McCracken}, {Marano},
  {Marinoni}, {Mazure}, {Meneux}, {Merighi}, {Pell{\`o}}, {Pollo}, {Radovich},
  {Vergani}, {Zucca}, {Romano}, {Grado}, \& {Limatola}}]{Lamareille2009}
{Lamareille}, F., {Brinchmann}, J., {Contini}, T., {et~al.} 2009, \aap, 495, 53

\bibitem[{{Lamastra} {et~al.}(2009){Lamastra}, {Bianchi}, {Matt}, {Perola},
  {Barcons}, \& {Carrera}}]{Lamastra2009}
{Lamastra}, A., {Bianchi}, S., {Matt}, G., {et~al.} 2009, \aap, 504, 73

\bibitem[{{Lanzuisi} {et~al.}(2017){Lanzuisi}, {Delvecchio}, {Berta}, {Brusa},
  {Comastri}, {Gilli}, {Gruppioni}, {Marchesi}, {Perna}, {Pozzi}, {Salvato},
  {Symeonidis}, {Vignali}, {Vito}, {Volonteri}, \& {Zamorani}}]{Lanzuisi2017}
{Lanzuisi}, G., {Delvecchio}, I., {Berta}, S., {et~al.} 2017, \aap, 602, A123

\bibitem[{{Lawrence} {et~al.}(2007){Lawrence}, {Warren}, {Almaini}, {Edge},
  {Hambly}, {Jameson}, {Lucas}, {Casali}, {Adamson}, {Dye}, {Emerson},
  {Foucaud}, {Hewett}, {Hirst}, {Hodgkin}, {Irwin}, {Lodieu}, {McMahon},
  {Simpson}, {Smail}, {Mortlock}, \& {Folger}}]{Lawrence2007}
{Lawrence}, A., {Warren}, S.~J., {Almaini}, O., {et~al.} 2007, \mnras, 379,
  1599

\bibitem[{{Le F{\`e}vre} {et~al.}(2013){Le F{\`e}vre}, {Cassata}, {Cucciati},
  {Garilli}, {Ilbert}, {Le Brun}, {Maccagni}, {Moreau}, {Scodeggio}, {Tresse},
  {Zamorani}, {Adami}, {Arnouts}, {Bardelli}, {Bolzonella}, {Bondi},
  {Bongiorno}, {Bottini}, {Cappi}, {Charlot}, {Ciliegi}, {Contini}, {de la
  Torre}, {Foucaud}, {Franzetti}, {Gavignaud}, {Guzzo}, {Iovino}, {Lemaux},
  {L{\'o}pez-Sanjuan}, {McCracken}, {Marano}, {Marinoni}, {Mazure}, {Mellier},
  {Merighi}, {Merluzzi}, {Paltani}, {Pell{\`o}}, {Pollo}, {Pozzetti},
  {Scaramella}, {Tasca}, {Vergani}, {Vettolani}, {Zanichelli}, \&
  {Zucca}}]{LeFevre2013}
{Le F{\`e}vre}, O., {Cassata}, P., {Cucciati}, O., {et~al.} 2013, \aap, 559,
  A14

\bibitem[{{Le F{\`e}vre} {et~al.}(2004{\natexlab{a}}){Le F{\`e}vre}, {Mellier},
  {McCracken}, {Foucaud}, {Gwyn}, {Radovich}, {Dantel-Fort}, {Bertin},
  {Moreau}, {Cuillandre}, {Pierre}, {Le Brun}, {Mazure}, \&
  {Tresse}}]{LeFevre2004}
{Le F{\`e}vre}, O., {Mellier}, Y., {McCracken}, H.~J., {et~al.}
  2004{\natexlab{a}}, \aap, 417, 839

\bibitem[{{Le F{\`e}vre} {et~al.}(2005){Le F{\`e}vre}, {Vettolani}, {Garilli},
  {Tresse}, {Bottini}, {Le Brun}, {Maccagni}, {Picat}, {Scaramella},
  {Scodeggio}, {Zanichelli}, {Adami}, {Arnaboldi}, {Arnouts}, {Bardelli},
  {Bolzonella}, {Cappi}, {Charlot}, {Ciliegi}, {Contini}, {Foucaud},
  {Franzetti}, {Gavignaud}, {Guzzo}, {Ilbert}, {Iovino}, {McCracken}, {Marano},
  {Marinoni}, {Mathez}, {Mazure}, {Meneux}, {Merighi}, {Paltani}, {Pell{\`o}},
  {Pollo}, {Pozzetti}, {Radovich}, {Zamorani}, {Zucca}, {Bondi}, {Bongiorno},
  {Busarello}, {Lamareille}, {Mellier}, {Merluzzi}, {Ripepi}, \&
  {Rizzo}}]{LeFevre2005}
{Le F{\`e}vre}, O., {Vettolani}, G., {Garilli}, B., {et~al.} 2005, \aap, 439,
  845

\bibitem[{{Le F{\`e}vre} {et~al.}(2004{\natexlab{b}}){Le F{\`e}vre},
  {Vettolani}, {Paltani}, {Tresse}, {Zamorani}, {Le Brun}, {Moreau}, {Bottini},
  {Maccagni}, {Picat}, {Scaramella}, {Scodeggio}, {Zanichelli}, {Adami},
  {Arnouts}, {Bardelli}, {Bolzonella}, {Cappi}, {Charlot}, {Contini},
  {Foucaud}, {Franzetti}, {Garilli}, {Gavignaud}, {Guzzo}, {Ilbert}, {Iovino},
  {McCracken}, {Mancini}, {Marano}, {Marinoni}, {Mathez}, {Mazure}, {Meneux},
  {Merighi}, {Pell{\`o}}, {Pollo}, {Pozzetti}, {Radovich}, {Zucca},
  {Arnaboldi}, {Bondi}, {Bongiorno}, {Busarello}, {Ciliegi}, {Gregorini},
  {Mellier}, {Merluzzi}, {Ripepi}, \& {Rizzo}}]{LeFevre2004b}
{Le F{\`e}vre}, O., {Vettolani}, G., {Paltani}, S., {et~al.}
  2004{\natexlab{b}}, \aap, 428, 1043

\bibitem[{{Leitherer} {et~al.}(2002){Leitherer}, {Li}, {Calzetti}, \&
  {Heckman}}]{Leitherer2002}
{Leitherer}, C., {Li}, I.~H., {Calzetti}, D., \& {Heckman}, T.~M. 2002, \apjs,
  140, 303

\bibitem[{{Leslie} {et~al.}(2016){Leslie}, {Kewley}, {Sanders}, \&
  {Lee}}]{Leslie2016}
{Leslie}, S.~K., {Kewley}, L.~J., {Sanders}, D.~B., \& {Lee}, N. 2016, \mnras,
  455, L82

\bibitem[{{Lonsdale} {et~al.}(2004){Lonsdale}, {Polletta}, {Surace}, {Shupe},
  {Fang}, {Xu}, {Smith}, {Siana}, {Rowan-Robinson}, {Babbedge}, {Oliver},
  {Pozzi}, {Davoodi}, {Owen}, {Padgett}, {Frayer}, {Jarrett}, {Masci},
  {O'Linger}, {Conrow}, {Farrah}, {Morrison}, {Gautier}, {Franceschini},
  {Berta}, {Perez-Fournon}, {Hatziminaoglou}, {Afonso-Luis}, {Dole}, {Stacey},
  {Serjeant}, {Pierre}, {Griffin}, \& {Puetter}}]{Lonsdale2004}
{Lonsdale}, C., {Polletta}, M. d.~C., {Surace}, J., {et~al.} 2004, \apjs, 154,
  54

\bibitem[{{Lonsdale} {et~al.}(2003){Lonsdale}, {Smith}, {Rowan-Robinson},
  {Surace}, {Shupe}, {Xu}, {Oliver}, {Padgett}, {Fang}, {Conrow},
  {Franceschini}, {Gautier}, {Griffin}, {Hacking}, {Masci}, {Morrison},
  {O'Linger}, {Owen}, {P{\'e}rez-Fournon}, {Pierre}, {Puetter}, {Stacey},
  {Castro}, {Polletta}, {Farrah}, {Jarrett}, {Frayer}, {Siana}, {Babbedge},
  {Dye}, {Fox}, {Gonzalez-Solares}, {Salaman}, {Berta}, {Condon}, {Dole}, \&
  {Serjeant}}]{Lonsdale2003}
{Lonsdale}, C.~J., {Smith}, H.~E., {Rowan-Robinson}, M., {et~al.} 2003, \pasp,
  115, 897

\bibitem[{{Lynden-Bell}(1969)}]{Lynden1969}
{Lynden-Bell}, D. 1969, \nat, 223, 690

\bibitem[{{Madau} {et~al.}(1996){Madau}, {Ferguson}, {Dickinson}, {Giavalisco},
  {Steidel}, \& {Fruchter}}]{Madau1996}
{Madau}, P., {Ferguson}, H.~C., {Dickinson}, M.~E., {et~al.} 1996, \mnras, 283,
  1388

\bibitem[{{Magorrian} {et~al.}(1998){Magorrian}, {Tremaine}, {Richstone},
  {Bender}, {Bower}, {Dressler}, {Faber}, {Gebhardt}, {Green}, {Grillmair},
  {Kormendy}, \& {Lauer}}]{Magorrian1998}
{Magorrian}, J., {Tremaine}, S., {Richstone}, D., {et~al.} 1998, \aj, 115, 2285

\bibitem[{{Manzoni} {et~al.}(2021){Manzoni}, {Scodeggio}, {Baugh}, {Norberg},
  {De Lucia}, {Fritz}, {Haines}, {Zamorani}, {Gargiulo}, {Guzzo}, {Iovino},
  {Ma{\l}ek}, {Pollo}, {Siudek}, \& {Vergani}}]{Manzoni2021}
{Manzoni}, G., {Scodeggio}, M., {Baugh}, C.~M., {et~al.} 2021, \na, 84, 101515

\bibitem[{{Maraston} {et~al.}(2013){Maraston}, {Pforr}, {Henriques}, {Thomas},
  {Wake}, {Brownstein}, {Capozzi}, {Tinker}, {Bundy}, {Skibba}, {Beifiori},
  {Nichol}, {Edmondson}, {Schneider}, {Chen}, {Masters}, {Steele}, {Bolton},
  {York}, {Weaver}, {Higgs}, {Bizyaev}, {Brewington}, {Malanushenko},
  {Malanushenko}, {Snedden}, {Oravetz}, {Pan}, {Shelden}, \&
  {Simmons}}]{Maraston2013}
{Maraston}, C., {Pforr}, J., {Henriques}, B.~M., {et~al.} 2013, \mnras, 435,
  2764

\bibitem[{{Masoura} {et~al.}(2021){Masoura}, {Mountrichas}, {Georgantopoulos},
  \& {Plionis}}]{Masoura2021}
{Masoura}, V.~A., {Mountrichas}, G., {Georgantopoulos}, I., \& {Plionis}, M.
  2021, \aap, 646, A167

\bibitem[{{Matsuoka} {et~al.}(2015){Matsuoka}, {Strauss}, {Shen}, {Brandt},
  {Greene}, {Ho}, {Schneider}, {Sun}, \& {Trump}}]{Matsuoka2015}
{Matsuoka}, Y., {Strauss}, M.~A., {Shen}, Y., {et~al.} 2015, \apj, 811, 91

\bibitem[{{McCracken} {et~al.}(2003){McCracken}, {Radovich}, {Bertin},
  {Mellier}, {Dantel-Fort}, {Le F{\`e}vre}, {Cuilland re}, {Gwyn}, {Foucaud},
  \& {Zamorani}}]{McCracken2003}
{McCracken}, H.~J., {Radovich}, M., {Bertin}, E., {et~al.} 2003, \aap, 410, 17

\bibitem[{{Menci} {et~al.}(2008){Menci}, {Fiore}, {Puccetti}, \&
  {Cavaliere}}]{Menci2008}
{Menci}, N., {Fiore}, F., {Puccetti}, S., \& {Cavaliere}, A. 2008, \apj, 686,
  219

\bibitem[{{Moutard} {et~al.}(2016){Moutard}, {Arnouts}, {Ilbert}, {Coupon},
  {Hudelot}, {Vibert}, {Comte}, {Conseil}, {Davidzon}, {Guzzo}, {Llebaria},
  {Martin}, {McCracken}, {Milliard}, {Morrison}, {Schiminovich}, {Treyer}, \&
  {Van Werbaeke}}]{Moutard2016}
{Moutard}, T., {Arnouts}, S., {Ilbert}, O., {et~al.} 2016, \aap, 590, A102

\bibitem[{{Mullaney} {et~al.}(2015){Mullaney}, {Alexander}, {Aird}, {Bernhard},
  {Daddi}, {Del Moro}, {Dickinson}, {Elbaz}, {Harrison}, {Juneau}, {Liu},
  {Pannella}, {Rosario}, {Santini}, {Sargent}, {Schreiber}, {Simpson}, \&
  {Stanley}}]{Mullaney2015}
{Mullaney}, J.~R., {Alexander}, D.~M., {Aird}, J., {et~al.} 2015, \mnras, 453,
  L83

\bibitem[{{Mullaney} {et~al.}(2013){Mullaney}, {Alexander}, {Fine}, {Goulding},
  {Harrison}, \& {Hickox}}]{Mullaney2013}
{Mullaney}, J.~R., {Alexander}, D.~M., {Fine}, S., {et~al.} 2013, \mnras, 433,
  622

\bibitem[{{Mullaney} {et~al.}(2012){Mullaney}, {Daddi}, {B{\'e}thermin},
  {Elbaz}, {Juneau}, {Pannella}, {Sargent}, {Alexander}, \&
  {Hickox}}]{Mullaney2012}
{Mullaney}, J.~R., {Daddi}, E., {B{\'e}thermin}, M., {et~al.} 2012, \apjl, 753,
  L30

\bibitem[{{Noll} {et~al.}(2009){Noll}, {Burgarella}, {Giovannoli}, {Buat},
  {Marcillac}, \& {Mu{\~n}oz-Mateos}}]{Noll2009}
{Noll}, S., {Burgarella}, D., {Giovannoli}, E., {et~al.} 2009, \aap, 507, 1793

\bibitem[{{Osterbrock} \& {Ferland}(2006)}]{Osterbrock2006}
{Osterbrock}, D.~E. \& {Ferland}, G.~J. 2006, {Astrophysics of gaseous nebulae
  and active galactic nuclei}

\bibitem[{{Page} {et~al.}(2012){Page}, {Symeonidis}, {Vieira}, {Altieri},
  {Amblard}, {Arumugam}, {Aussel}, {Babbedge}, {Blain}, {Bock}, {Boselli},
  {Buat}, {Castro-Rodr{\'\i}guez}, {Cava}, {Chanial}, {Clements}, {Conley},
  {Conversi}, {Cooray}, {Dowell}, {Dubois}, {Dunlop}, {Dwek}, {Dye}, {Eales},
  {Elbaz}, {Farrah}, {Fox}, {Franceschini}, {Gear}, {Glenn}, {Griffin},
  {Halpern}, {Hatziminaoglou}, {Ibar}, {Isaak}, {Ivison}, {Lagache},
  {Levenson}, {Lu}, {Madden}, {Maffei}, {Mainetti}, {Marchetti}, {Nguyen},
  {O'Halloran}, {Oliver}, {Omont}, {Panuzzo}, {Papageorgiou}, {Pearson},
  {P{\'e}rez-Fournon}, {Pohlen}, {Rawlings}, {Rigopoulou}, {Riguccini},
  {Rizzo}, {Rodighiero}, {Roseboom}, {Rowan-Robinson}, {Portal}, {Schulz},
  {Scott}, {Seymour}, {Shupe}, {Smith}, {Stevens}, {Trichas}, {Tugwell},
  {Vaccari}, {Valtchanov}, {Viero}, {Vigroux}, {Wang}, {Ward}, {Wright}, {Xu},
  \& {Zemcov}}]{Page2012}
{Page}, M.~J., {Symeonidis}, M., {Vieira}, J.~D., {et~al.} 2012, \nat, 485, 213

\bibitem[{{P{\'e}rez-Gonz{\'a}lez} {et~al.}(2008){P{\'e}rez-Gonz{\'a}lez},
  {Rieke}, {Villar}, {et~al.}}]{Perez2008}
{P{\'e}rez-Gonz{\'a}lez}, P.~G., {Rieke}, G.~H., {Villar}, V., {et~al.} 2008,
  \apj, 675, 234

\bibitem[{{Polletta} {et~al.}(2011){Polletta}, {Nesvadba}, {Neri}, {Omont},
  {Berta}, \& {Bergeron}}]{Polletta2011}
{Polletta}, M., {Nesvadba}, N.~P.~H., {Neri}, R., {et~al.} 2011, \aap, 533, A20

\bibitem[{{Rodighiero} {et~al.}(2011){Rodighiero}, {Daddi}, {Baronchelli},
  {Cimatti}, {Renzini}, {Aussel}, {Popesso}, {Lutz}, {Andreani}, {Berta},
  {Cava}, {Elbaz}, {Feltre}, {Fontana}, {F{\"o}rster Schreiber},
  {Franceschini}, {Genzel}, {Grazian}, {Gruppioni}, {Ilbert}, {Le Floch},
  {Magdis}, {Magliocchetti}, {Magnelli}, {Maiolino}, {McCracken}, {Nordon},
  {Poglitsch}, {Santini}, {Pozzi}, {Riguccini}, {Tacconi}, {Wuyts}, \&
  {Zamorani}}]{Rodighiero2011}
{Rodighiero}, G., {Daddi}, E., {Baronchelli}, I., {et~al.} 2011, \apjl, 739,
  L40

\bibitem[{{Rola} {et~al.}(1997){Rola}, {Terlevich}, \& {Terlevich}}]{Rola1997}
{Rola}, C.~S., {Terlevich}, E., \& {Terlevich}, R.~J. 1997, \mnras, 289, 419

\bibitem[{{Rosa-Gonz{\'a}lez} {et~al.}(2002){Rosa-Gonz{\'a}lez}, {Terlevich},
  \& {Terlevich}}]{RosaGonzalez2002}
{Rosa-Gonz{\'a}lez}, D., {Terlevich}, E., \& {Terlevich}, R. 2002, \mnras, 332,
  283

\bibitem[{{Rosario} {et~al.}(2012){Rosario}, {Santini}, {Lutz}, {Shao},
  {Maiolino}, {Alexander}, {Altieri}, {Andreani}, {Aussel}, {Bauer}, {Berta},
  {Bongiovanni}, {Brandt}, {Brusa}, {Cepa}, {Cimatti}, {Cox}, {Daddi}, {Elbaz},
  {Fontana}, {F{\"o}rster Schreiber}, {Genzel}, {Grazian}, {Le Floch},
  {Magnelli}, {Mainieri}, {Netzer}, {Nordon}, {P{\'e}rez Garcia}, {Poglitsch},
  {Popesso}, {Pozzi}, {Riguccini}, {Rodighiero}, {Salvato}, {Sanchez-Portal},
  {Sturm}, {Tacconi}, {Valtchanov}, \& {Wuyts}}]{Rosario2012}
{Rosario}, D.~J., {Santini}, P., {Lutz}, D., {et~al.} 2012, \aap, 545, A45

\bibitem[{{Santini} {et~al.}(2012){Santini}, {Rosario}, {Shao}, {Lutz},
  {Maiolino}, {Alexander}, {Altieri}, {Andreani}, {Aussel}, {Bauer}, {Berta},
  {Bongiovanni}, {Brandt}, {Brusa}, {Cepa}, {Cimatti}, {Daddi}, {Elbaz},
  {Fontana}, {F{\"o}rster Schreiber}, {Genzel}, {Grazian}, {Le Floc'h},
  {Magnelli}, {Mainieri}, {Nordon}, {P{\'e}rez Garcia}, {Poglitsch}, {Popesso},
  {Pozzi}, {Riguccini}, {Rodighiero}, {Salvato}, {Sanchez-Portal}, {Sturm},
  {Tacconi}, {Valtchanov}, \& {Wuyts}}]{Santini2012}
{Santini}, P., {Rosario}, D.~J., {Shao}, L., {et~al.} 2012, \aap, 540, A109

\bibitem[{{Schaye} {et~al.}(2015){Schaye}, {Crain}, {Bower}, {Furlong},
  {Schaller}, {Theuns}, {Dalla Vecchia}, {Frenk}, {McCarthy}, {Helly},
  {Jenkins}, {Rosas-Guevara}, {White}, {Baes}, {Booth}, {Camps}, {Navarro},
  {Qu}, {Rahmati}, {Sawala}, {Thomas}, \& {Trayford}}]{Schaye2015}
{Schaye}, J., {Crain}, R.~A., {Bower}, R.~G., {et~al.} 2015, \mnras, 446, 521

\bibitem[{{Schreiber} {et~al.}(2015){Schreiber}, {Pannella}, {Elbaz},
  {B{\'e}thermin}, {Inami}, {Dickinson}, {Magnelli}, {Wang}, {Aussel}, {Daddi},
  {Juneau}, {Shu}, {Sargent}, {Buat}, {Faber}, {Ferguson}, {Giavalisco},
  {Koekemoer}, {Magdis}, {Morrison}, {Papovich}, {Santini}, \&
  {Scott}}]{Schreiber2015}
{Schreiber}, C., {Pannella}, M., {Elbaz}, D., {et~al.} 2015, \aap, 575, A74

\bibitem[{{Schwarz}(1978)}]{Schwarz1978}
{Schwarz}, U.~J. 1978, \aap, 65, 345

\bibitem[{{Scodeggio} {et~al.}(2018){Scodeggio}, {Guzzo}, {Garilli}, {Granett},
  {Bolzonella}, {de la Torre}, {Abbas}, {Adami}, {Arnouts}, {Bottini}, {Cappi},
  {Coupon}, {Cucciati}, {Davidzon}, {Franzetti}, {Fritz}, {Iovino}, {Krywult},
  {Le Brun}, {Le F{\`e}vre}, {Maccagni}, {Ma{\l}ek}, {Marchetti}, {Marulli},
  {Polletta}, {Pollo}, {Tasca}, {Tojeiro}, {Vergani}, {Zanichelli}, {Bel},
  {Branchini}, {De Lucia}, {Ilbert}, {McCracken}, {Moutard}, {Peacock},
  {Zamorani}, {Burden}, {Fumana}, {Jullo}, {Marinoni}, {Mellier}, {Moscardini},
  \& {Percival}}]{Scodeggio2018}
{Scodeggio}, M., {Guzzo}, L., {Garilli}, B., {et~al.} 2018, \aap, 609, A84

\bibitem[{{Shao} {et~al.}(2010){Shao}, {Lutz}, {Nordon}, {Maiolino},
  {Alexander}, {Altieri}, {Andreani}, {Aussel}, {Bauer}, {Berta},
  {Bongiovanni}, {Brandt}, {Brusa}, {Cava}, {Cepa}, {Cimatti}, {Daddi},
  {Dominguez-Sanchez}, {Elbaz}, {F{\"o}rster Schreiber}, {Geis}, {Genzel},
  {Grazian}, {Gruppioni}, {Magdis}, {Magnelli}, {Mainieri}, {P{\'e}rez
  Garc{\'\i}a}, {Poglitsch}, {Popesso}, {Pozzi}, {Riguccini}, {Rodighiero},
  {Rovilos}, {Saintonge}, {Salvato}, {Sanchez Portal}, {Santini}, {Sturm},
  {Tacconi}, {Valtchanov}, {Wetzstein}, \& {Wieprecht}}]{Shao2010}
{Shao}, L., {Lutz}, D., {Nordon}, R., {et~al.} 2010, \aap, 518, L26

\bibitem[{{Shimizu} {et~al.}(2015){Shimizu}, {Mushotzky}, {Mel{\'e}ndez},
  {Koss}, \& {Rosario}}]{Shimizu2015}
{Shimizu}, T.~T., {Mushotzky}, R.~F., {Mel{\'e}ndez}, M., {Koss}, M., \&
  {Rosario}, D.~J. 2015, \mnras, 452, 1841

\bibitem[{{Shimizu} {et~al.}(2017){Shimizu}, {Mushotzky}, {Mel{\'e}ndez},
  {Koss}, {Barger}, \& {Cowie}}]{Shimizu2017}
{Shimizu}, T.~T., {Mushotzky}, R.~F., {Mel{\'e}ndez}, M., {et~al.} 2017,
  \mnras, 466, 3161

\bibitem[{{Silk} \& {Rees}(1998)}]{Silk1998}
{Silk}, J. \& {Rees}, M.~J. 1998, \aap, 331, L1

\bibitem[{{Silverman} {et~al.}(2009){Silverman}, {Lamareille}, {Maier},
  {Lilly}, {Mainieri}, {Brusa}, {Cappelluti}, {Hasinger}, {Zamorani},
  {Scodeggio}, {Bolzonella}, {Contini}, {Carollo}, {Jahnke}, {Kneib}, {Le
  F{\`e}vre}, {Merloni}, {Bardelli}, {Bongiorno}, {Brunner}, {Caputi},
  {Civano}, {Comastri}, {Coppa}, {Cucciati}, {de la Torre}, {de Ravel},
  {Elvis}, {Finoguenov}, {Fiore}, {Franzetti}, {Garilli}, {Gilli}, {Iovino},
  {Kampczyk}, {Knobel}, {Kova{\v{c}}}, {Le Borgne}, {Le Brun}, {Mignoli},
  {Pello}, {Peng}, {Perez Montero}, {Ricciardelli}, {Tanaka}, {Tasca},
  {Tresse}, {Vergani}, {Vignali}, {Zucca}, {Bottini}, {Cappi}, {Cassata},
  {Fumana}, {Griffiths}, {Kartaltepe}, {Koekemoer}, {Marinoni}, {McCracken},
  {Memeo}, {Meneux}, {Oesch}, {Porciani}, \& {Salvato}}]{Silverman2009}
{Silverman}, J.~D., {Lamareille}, F., {Maier}, C., {et~al.} 2009, \apj, 696,
  396

\bibitem[{{Speagle} {et~al.}(2014){Speagle}, {Steinhardt}, {Capak}, \&
  {Silverman}}]{Speagle2014}
{Speagle}, J.~S., {Steinhardt}, C.~L., {Capak}, P.~L., \& {Silverman}, J.~D.
  2014, \apjs, 214, 15

\bibitem[{{Stanley} {et~al.}(2017){Stanley}, {Alexander}, {Harrison},
  {Rosario}, {Wang}, {Aird}, {Bourne}, {Dunne}, {Dye}, {Eales}, {Knudsen},
  {Micha{\l}owski}, {Valiante}, {De Zotti}, {Furlanetto}, {Ivison}, {Maddox},
  \& {Smith}}]{Stanley2017}
{Stanley}, F., {Alexander}, D.~M., {Harrison}, C.~M., {et~al.} 2017, \mnras,
  472, 2221

\bibitem[{{Stanley} {et~al.}(2015){Stanley}, {Harrison}, {Alexander},
  {Swinbank}, {Aird}, {Del Moro}, {Hickox}, \& {Mullaney}}]{Stanley2015}
{Stanley}, F., {Harrison}, C.~M., {Alexander}, D.~M., {et~al.} 2015, \mnras,
  453, 591

\bibitem[{{Stemo} {et~al.}(2020){Stemo}, {Comerford}, {Barrows}, {Stern},
  {Assef}, \& {Griffith}}]{Stemo2020}
{Stemo}, A., {Comerford}, J.~M., {Barrows}, R.~S., {et~al.} 2020, \apj, 888, 78

\bibitem[{{Strateva} {et~al.}(2001){Strateva}, {Ivezi{\'c}}, {Knapp},
  {Narayanan}, {Strauss}, {Gunn}, {Lupton}, {Schlegel}, {Bahcall}, {Brinkmann},
  {Brunner}, {Budav{\'a}ri}, {Csabai}, {Castander}, {Doi}, {Fukugita},
  {Gy{\H{o}}ry}, {Hamabe}, {Hennessy}, {Ichikawa}, {Kunszt}, {Lamb}, {McKay},
  {Okamura}, {Racusin}, {Sekiguchi}, {Schneider}, {Shimasaku}, \&
  {York}}]{Strateva2001}
{Strateva}, I., {Ivezi{\'c}}, {\v{Z}}., {Knapp}, G.~R., {et~al.} 2001, \aj,
  122, 1861

\bibitem[{{Thomas} {et~al.}(2013){Thomas}, {Steele}, {Maraston}, {Johansson},
  {Beifiori}, {Pforr}, {Str{\"o}mb{\"a}ck}, {Tremonti}, {Wake}, {Bizyaev},
  {Bolton}, {Brewington}, {Brownstein}, {Comparat}, {Kneib}, {Malanushenko},
  {Malanushenko}, {Oravetz}, {Pan}, {Parejko}, {Schneider}, {Shelden},
  {Simmons}, {Snedden}, {Tanaka}, {Weaver}, \& {Yan}}]{Thomas2013}
{Thomas}, D., {Steele}, O., {Maraston}, C., {et~al.} 2013, \mnras, 431, 1383

\bibitem[{{Tombesi} {et~al.}(2010){Tombesi}, {Cappi}, {Reeves}, {Palumbo},
  {Yaqoob}, {Braito}, \& {Dadina}}]{Tombesi2010}
{Tombesi}, F., {Cappi}, M., {Reeves}, J.~N., {et~al.} 2010, \aap, 521, A57

\bibitem[{{Urry} \& {Padovani}(2000)}]{Urry2000}
{Urry}, M. \& {Padovani}, P. 2000, \pasp, 112, 1516

\bibitem[{{Vazdekis} {et~al.}(2010){Vazdekis}, {S{\'a}nchez-Bl{\'a}zquez},
  {Falc{\'o}n-Barroso}, {Cenarro}, {Beasley}, {Cardiel}, {Gorgas}, \&
  {Peletier}}]{Vazdekis2010}
{Vazdekis}, A., {S{\'a}nchez-Bl{\'a}zquez}, P., {Falc{\'o}n-Barroso}, J.,
  {et~al.} 2010, \mnras, 404, 1639

\bibitem[{{Vietri} {et~al.}(2020){Vietri}, {Mainieri}, {Kakkad}, {Netzer},
  {Perna}, {Circosta}, {Harrison}, {Zappacosta}, {Husemann}, {Padovani},
  {Bischetti}, {Bongiorno}, {Brusa}, {Carniani}, {Cicone}, {Comastri},
  {Cresci}, {Feruglio}, {Fiore}, {Lanzuisi}, {Mannucci}, {Marconi},
  {Piconcelli}, {Puglisi}, {Salvato}, {Schramm}, {Schulze}, {Scholtz},
  {Vignali}, \& {Zamorani}}]{Vietri2020}
{Vietri}, G., {Mainieri}, V., {Kakkad}, D., {et~al.} 2020, \aap, 644, A175

\bibitem[{{Whitaker} {et~al.}(2015){Whitaker}, {Franx}, {Bezanson}, {Brammer},
  {van Dokkum}, {Kriek}, {Labb{\'e}}, {Leja}, {Momcheva}, {Nelson}, {Rigby},
  {Rix}, {Skelton}, {van der Wel}, \& {Wuyts}}]{Whitaker2015}
{Whitaker}, K.~E., {Franx}, M., {Bezanson}, R., {et~al.} 2015, \apjl, 811, L12

\bibitem[{{Whitaker} {et~al.}(2014){Whitaker}, {Franx}, {Leja}, {van Dokkum},
  {Henry}, {Skelton}, {Fumagalli}, {Momcheva}, {Brammer}, {Labb{\'e}},
  {Nelson}, \& {Rigby}}]{Whitaker2014}
{Whitaker}, K.~E., {Franx}, M., {Leja}, J., {et~al.} 2014, \apj, 795, 104

\bibitem[{{Whitaker} {et~al.}(2012){Whitaker}, {van Dokkum}, {Brammer}, \&
  {Franx}}]{Whitaker2012}
{Whitaker}, K.~E., {van Dokkum}, P.~G., {Brammer}, G., \& {Franx}, M. 2012,
  \apjl, 754, L29

\bibitem[{{Woo} {et~al.}(2016){Woo}, {Bae}, {Son}, \& {Karouzos}}]{Woo2016}
{Woo}, J.-H., {Bae}, H.-J., {Son}, D., \& {Karouzos}, M. 2016, \apj, 817, 108

\bibitem[{{Wright} {et~al.}(2010){Wright}, {Eisenhardt}, {Mainzer}, {Ressler},
  {Cutri}, {Jarrett}, {Kirkpatrick}, {Padgett}, {McMillan}, {Skrutskie},
  {Stanford}, {Cohen}, {Walker}, {Mather}, {Leisawitz}, {Gautier}, {McLean},
  {Benford}, {Lonsdale}, {Blain}, {Mendez}, {Irace}, {Duval}, {Liu}, {Royer},
  {Heinrichsen}, {Howard}, {Shannon}, {Kendall}, {Walsh}, {Larsen}, {Cardon},
  {Schick}, {Schwalm}, {Abid}, {Fabinsky}, {Naes}, \& {Tsai}}]{Wright2010}
{Wright}, E.~L., {Eisenhardt}, P. R.~M., {Mainzer}, A.~K., {et~al.} 2010, \aj,
  140, 1868

\bibitem[{{Yang} {et~al.}(2017){Yang}, {Chen}, {Vito}, {Brandt}, {Alexander},
  {Luo}, {Sun}, {Xue}, {Bauer}, {Koekemoer}, {Lehmer}, {Liu}, {Schneider},
  {Shemmer}, {Trump}, {Vignali}, \& {Wang}}]{Yang2017}
{Yang}, G., {Chen}, C. T.~J., {Vito}, F., {et~al.} 2017, \apj, 842, 72

\bibitem[{{Zakamska} {et~al.}(2003){Zakamska}, {Strauss}, {Krolik}, {Collinge},
  {Hall}, {Hao}, {Heckman}, {Ivezi{\'c}}, {Richards}, {Schlegel}, {Schneider},
  {Strateva}, {Vanden Berk}, {Anderson}, \& {Brinkmann}}]{Zakamska2003}
{Zakamska}, N.~L., {Strauss}, M.~A., {Krolik}, J.~H., {et~al.} 2003, \aj, 126,
  2125

\bibitem[{{Zhuang} \& {Ho}(2019)}]{Zhuang2019}
{Zhuang}, M.-Y. \& {Ho}, L.~C. 2019, \apj, 882, 89

\bibitem[{{Zhuang} \& {Ho}(2020)}]{Zhuang2020}
{Zhuang}, M.-Y. \& {Ho}, L.~C. 2020, \apj, 896, 108

\bibitem[{{Zhuang} {et~al.}(2018){Zhuang}, {Ho}, \& {Shangguan}}]{Zhuang2018}
{Zhuang}, M.-Y., {Ho}, L.~C., \& {Shangguan}, J. 2018, \apj, 862, 118

\end{thebibliography}

\end{document}